\documentclass[review]{elsarticle}

\usepackage[dvipsnames]{xcolor}
\usepackage{lineno,hyperref}
\usepackage{amssymb}
\usepackage{amsmath}
\usepackage{xspace}

\newcommand{\de}[1]{{\rm d}#1}
\newcommand{\as}{\ensuremath{\alpha_{\rm s}}}
\newcommand{\asbar}{\ensuremath{\bar{\alpha}_{\rm s}}}
\newcommand{\aem}{\ensuremath{\alpha_{\rm em}}}
\newcommand{\RpA}{\ensuremath{R_{\rm p \it A}}}
\newcommand{\NC}{\ensuremath{N_{\rm C}}}
\newcommand{\Qs}{\ensuremath{Q_{\rm s0}}}
\newcommand{\QsA}{\ensuremath{Q_{\rm s0,\it A}}}
\newcommand{\Tp}{\ensuremath{T_{\rm p}}}
\newcommand{\TA}{\ensuremath{T_{A}}}

\begin{document}

\begin{frontmatter}

\title{
Probing nuclear structure with the Balitsky-Kovchegov equation in full impact-parameter dependence
}

\author[CVUT]{J. Cepila}
\author[CVUT]{M. Matas}
\author[CVUT]{M. Vaculciak}
\address[CVUT]{Faculty of Nuclear Sciences and Physical Engineering, Czech Technical University in Prague, Czech Republic}

\begin{abstract}
Building on the newly available solution of the Balitsky-Kovchegov (BK) equation with the full impact-parameter dependence, we extend the study of parton evolution from proton to nuclear targets.
Since a key part of the scientific programme for future experimental facilities such as the EIC is to study gluon dynamics and shed new light on the phenomenon of parton saturation, we present predictions for key processes, such as deep-inelastic scattering or the diffractive production of vector mesons, on a variety of nuclear targets.
Besides the standard BK equation, we employ its linearised version to identify a promising channel to search for gluon saturation in the nuclear collisions.
Furthermore, we implement a tetrahedral model of oxygen to search for deviations from the standard, isotropic, Woods-Saxon approach.
In addition to the future colliders, the presented results are also of interest for the current studies of nuclear vector meson production at the LHC.
\end{abstract}


\end{frontmatter}

\section{Introduction \label{sec:intro}}

The structure of nuclei, specifically its gluon component, has been studied extensively in the perturbative regime of QCD. Besides nuclear deep-inelastic scattering, the coherent photoproduction of vector mesons was shown a long time ago to be a sensitive probe of the nuclear colour field~\cite{Ryskin:1992ui, Brodsky:1994kf}. Such processes have been studied in the past and current experimental facilities, such as HERA~\cite{Ivanov:2004ax, Aaron:2009aa, Newman:2013ada} and LHC~\cite{Baltz:2007kq, Contreras:2015dqa, Klein:2019qfb, ALICE:2023jgu} and are of great interest for future \mbox{electron-ion} facilities such as EIC~\cite{Accardi:2012qut, Achenbach:2023pba} or LHeC~\cite{LHeCStudyGroup:2012zhm}. Both the recent measurements and those to come have renewed interest in predictions of such \mbox{nuclei-induced} processes~\cite{Bendova:2020hbb, Mantysaari:2025ltq, Goncalves:2025wwt}.

Compared to previous facilities, EIC will, for the first time, provide electron-ion collisions over a larger range of the virtuality--Bjorken-$x$ plane~\cite{Achenbach:2023pba}. The kinematic region to be explored is important for studying parton saturation, a phenomenon necessary for a consistent description of hadron structure. 
In addition to extending the kinematic region, new data should be available for a variety of light and heavy nuclear targets~\cite{AbdulKhalek:2021gbh}.

These new data will shed light on the difference between the gluon structure of nuclear and proton targets. At low Bjorken-$x$, the nuclear structure function per nucleon is smaller than that of the proton. This suppression, known as shadowing, has been interpreted in phenomenological studies as a consequence of multiple scattering~\cite{Armesto:2006ph, Frankfurt:2011cs}. Additional contributions may arise from gluon recombination due to the overlap of gluon fields from different nucleons~\cite{Gribov:1984tu, Mueller:1985wy}. At a certain scale, the gluon recombination balances splitting and the gluon density ceases to grow with increasing interaction energy. Such a regime is referred to as gluon saturation. Quantitatively, the evolution of the gluon density in a fast-moving frame is described by non-linear evolution equations. One of the most commonly used evolution equations in this context is the Balitsky-Kovchegov (BK) equation~\cite{Balitsky:1995ub, Kovchegov:1999yj}, which has been used to describe phenomena observed in proton collisions. This has been done using the impact-parameter-independent framework (e.g. Ref.~\cite{Albacete:2010sy}), and was later extended to a framework including the dependence on the impact-parameter magnitude~(\cite{Cepila:2018faq, Bendova:2019psy}) and even the relative angle between the dipole size and impact parameter (\cite{Berger:2011ew, Cepila:2023pvh, Cepila:2025rkn}). The BK evolution equation has also been used to evolve the gluon structure of the nucleus in Refs.~\cite{Bendova:2020hbb, Cepila:2020xol}, and compared to the more commonly used approach via the Glauber model in the framework with explicit dependence on the magnitude of the impact parameter. 

The work presented in this paper builds on the recently published solution of the target-rapidity BK equation with the full impact parameter dependence for the proton target, presented in Refs.~\cite{Cepila:2023pvh, Cepila:2025rkn}, and extends it to nuclear targets according to the procedure described in Refs.~\cite{Bendova:2020hbb, Cepila:2020xol}. Our solution provides theoretical predictions for processes connected to saturation physics, to be measured at the EIC and other future experimental facilities. We chose a palette of nuclear targets of potential interest for EIC, namely C, O, Ca, Fe, Cu, Au and Pb. Since oxygen–oxygen and proton–oxygen collisions are on the current LHC agenda, and since various models exist for describing the nuclear structure, we also adopted an $\alpha$-cluster approach for the oxygen profile alongside the standard Woods–Saxon distribution. This model is based on Refs.~\cite{Li:2020vrg, Behera:2021zhi}, and we provide its implementation as a Python PyPI library~\cite{nuclear_helper}.

The results of the nuclear BK equation are used to predict the nuclear structure functions, the nuclear modification factor $\RpA$, and the cross-sections of $J/\psi$ vector meson photoproduction. In all cases, good agreement with available experimental data is found.

To study the onset of saturation effects in nuclear targets in a quantitative way, we introduce a suppression of the non-linear term in the BK evolution equation. In schematic form, the BK equation reads $ \partial_\eta N = K \otimes (N-\kappa N^2)$, where the second term on the r.h.s., proportional to $N^2$, encodes the non-linear dynamics. The scalar regulator $\kappa\in (0,1)$ controls the strength of the saturation effects such that in the limiting cases one either recovers the original BK equation ($\kappa=1$), or gets its linearised version ($\kappa=0$), equivalent to the BFKL equation~\cite{Kuraev:1977fs, Balitsky:1978ic, Lipatov:1976zz, Mueller:1993rr}. 

By comparing theoretical predictions from the standard and linearised BK equations, we identify nuclear vector meson production, and especially its differential cross section, as a particularly promising channel to probe saturation effects.

The work is organised as follows: Sec.~\ref{sec:overview} presents an overview of the BK formalism and its extension to nuclear targets, together with a description of the approach to modelling the tetrahedral oxygen target. In Sec.~\ref{sec:solutions}, we compare the dipole amplitudes obtained from the standard and linearised BK equation for proton and lead targets, Sec.~\ref{sec:sf} presents the predictions for the nuclear structure functions and the vector meson production cross-sections and show the comparison to data, and Sec.~\ref{sec:summary} briefly summarises the presented work.

\section{Formalism overview\label{sec:overview}}

The Balitsky-Kovchegov equation~\cite{Balitsky:1995ub, Kovchegov:1999yj, Kovchegov_2000, Balitsky_2007, Kovchegov_2007}
\begin{eqnarray}
\frac{\partial N(\vec r,\vec b,\eta)}{\partial \eta}&=&\int d \vec{r}_1 K(\vec r,\vec r_1,\vec r_2)\big(N(\vec r_1,\vec b_1,\eta_1)+N(\vec r_2,\vec b_2,\eta_2)\nonumber\\
&&-N(\vec r,\vec b,\eta)-\kappa N(\vec r_1,\vec b_1,\eta_1)N(\vec r_2,\vec b_2,\eta_2)\big). 
\end{eqnarray}
provides a phenomenological tool to incorporate gluon saturation within the colour dipole picture. It evolves $N(\vec{r}, \vec{b},\eta)$, the amplitude for the interaction between the colour field of the target and a quark-antiquark dipole of size~$\vec{r}$ and impact parameter~$\vec{b}$, with respect to $\eta$, the rapidity of the target. This relates to the Bjorken-$x$ as \mbox{$\eta = \ln (x_0/x)$} with $x_0 = 0.01$. Both the dipole size and the impact parameter are two-dimensional vectors in the light-cone transverse plane with sizes $r \equiv |\vec{r}|$ and $b \equiv |\vec{b}|$. The terms on the r.h.s of the equation correspond to the parent dipole $(\vec r,\vec b,\eta)$ and two daughter dipoles $(\vec r_1,\vec b_1,\eta_1)$ and $(\vec r_2,\vec b_2,\eta_2)$. The rapidities $\eta_1,\eta_2$ introduce a non-locality to the equation via 
\begin{equation}
\eta_i=\eta - \mathrm{max}\left\lbrace 0, \ln\left(\frac{r^2}{r_i^2}\right)\right\rbrace.
\end{equation}
The kernel $K$ in the collinearly-improved target-rapidity framework reads~\cite{Ducloue:2019ezk}
\begin{equation}
    K(\vec r,\vec r_1,\vec r_2)=\frac{\asbar}{2\pi}\frac{r^2}{r_1^2 r_2^2}\left( \frac{r^2}{\mathrm{min}(r_1^2,r_2^2)}\right)^{\pm\asbar A_1},
\end{equation}
where the constant $A_1 = 11/12$. The sign in exponent is positive whenever $r<\mathrm{min}(r_1,r_2)$ and negative otherwise. The coupling constant is defined as $\asbar =(\NC/\pi)\as$ with $\as = \as(\mathrm{min}(r,r_1,r_2))$ the running coupling constant evaluated in the variable-number-of-ﬂavours scheme, see Ref.~\cite{Bendova:2019psy}. The free parameter of the coupling constant is the infrared regulator $C$, and the number of colours is $\NC=3$.

The solution of (target-rapidity, collinearly improved) BK equation in various levels of approximation has been demonstrated to be a viable tool for describing experimental data for various processes (such as DIS, vector meson production, DVCS, etc.) off a proton target~\cite{Ducloue:2019ezk, Bendova:2019psy, Bendova:2022xhw}. This equation has recently been solved, including the dependence on three out of the four total spatial degrees of freedom, i.e. dependence on the dipole size $r$, impact parameter size $b$, and their relative angle $\theta$, using  initial conditions of the form 
\begin{equation}
    N(r,b,\theta,\eta=0)=1-\exp\left(-\frac{1}{4}(\Qs^2r^2)^\gamma \Tp(b,r)(1+c\cos(2\theta))\right)
\end{equation}
with
\begin{equation}
    \Tp(b,r)=\exp\left(-\frac{b^2+(r/2)^2}{2B}\right).
\end{equation}
Here $\Qs^2,\gamma,c, B$ are free parameters chosen to describe available experimental data for proton-induced processes such as DIS or production of various vector mesons, as described in more detail in  Refs.~\cite{Cepila:2023pvh, Cepila:2025rkn}. The parameter values in Table~\ref{tab:1} differ from those in Refs.~\cite{Cepila:2023pvh, Cepila:2025rkn}, as additional experimental data have been used for their extraction, namely DVCS and the electro-production of various vector mesons off a proton target.

Due to the computational complexity of the 3D BK equation, the parameters in Table~1 do not come from a full fitting procedure, but rather from an estimate based on a coarse scanning of the parameter space. The palette of experimental measurements used for this estimate consists of proton structure functions~\cite{Aaron:2009aa}, differential and total cross-sections of vector-meson production ($\phi$~\cite{Aaron:2009xp,ZEUS:2005bhf}, $\rho$~\cite{H1:2020lzc,CMS:2019awk,Breitweg:1997ed,Chekanov:2007zr,Aaron:2009xp}, $\omega$~\cite{ZEUS:1996zse,ZEUS:2000swq}, $J/\psi$~\cite{Aaron:2009xp,ZEUS:2005bhf,Alexa:2013xxa,LHCb:2018rcm,ALICE:2014eof,Aktas:2005xu,Chekanov:2002xi}, $\psi(2S)$~\cite{LHCb:2018rcm,Aaij:2014iea}, and $\Upsilon(1S)$~\cite{Aaij:2015kea,CMS:2018bbk,Breitweg:1998ki,Chekanov:2009zz,Adloff:2000vm}) and DVCS~\cite{H1:2009wnw}, all of them in virtuality range $Q^2 \in [0, 35]~{\rm GeV}^2$, resulting in $\chi^2/ndof = 6.4$. While this $\chi^2$ may seem large, it should be noted that this study is simultaneously taking into account a large variety of experimental observables rather than a narrow set of data for one particular process. Given that several of the included data sets have very narrow experimental uncertainties, while the present framework is expected to carry at least a few-percent systematic theory uncertainty (e.g., missing higher-order effects), a $\chi^2$ close to unity is not expected; substantially smaller values would likely indicate that theory uncertainties are being effectively absorbed into the fit rather than genuinely predicted. Furthermore, no K-factor has been introduced to improve the data description.
\begin{table}[!t]
    \centering
    \caption{Parameters of the solution to the BK including explicit dependence on ($r,b,\theta$) according to~\cite{Cepila:2025rkn}. Parameters were obtained for quark masses $m_{\rm uds} = 0.1~\mathrm{GeV}$, $m_{\rm c} = 1.27~\mathrm{GeV}$ and $m_{b} = 4.18~\mathrm{GeV}$.
    }
    \begin{tabular}{ccccc}
    $\Qs^2$ & $\gamma$ & $c$ & $B$ & $C$ \\\hline
    0.496 GeV$^{2}$ & 1.25 & 1 & 3.4 GeV$^{2}$ & 5 
    \end{tabular}
    \label{tab:1}
\end{table}

\subsection{Nuclear BK equation}
As this newly obtained solution was only validated against available experimental data probing the structure of the proton, we aim to extend the palette of phenomenological predictions to processes including nuclear targets. The transition from the proton BK equation to the nuclear one follows the approach of Refs.~\cite{Bendova:2020hbb, Cepila:2020xol} and consists of two modifications of the initial condition. First, the $A$-dependent form of saturation scale is used
\begin{equation}
    \Qs^2 \rightarrow \QsA^2=0.1\mathrm{~GeV}^2 A^{1/3},
\end{equation}
where $A$ is the nucleus mass number, and second, the profile function, modelling the proton as a Gaussian, is replaced with the nuclear thickness functions, normalised such that
\begin{equation}
    \Tp(b,r)\rightarrow \TA(b,r)=\frac{T(\sqrt{b^2+r^2/4})}{\max \big( T(\sqrt{b^2+r^2/4})\big)}.
\end{equation}
In most cases, the nuclear profile is modelled according to the Woods-Saxon (2-point Fermi) distribution for the corresponding nucleus~\cite{DEVRIES1987495} 
\begin{equation}
    T\Big(\tilde b = \sqrt{b^2 + r^2/4} \Big)=\int\limits_{-\infty}^{+\infty} \mathrm{d}z \frac{\rho_0}{1+\exp\Bigg[\Big(\sqrt{\tilde b^2+z^2}-R\Big)/a\Bigg]},
\end{equation}
where $\max  (T(\tilde b)) = T(0)$. For very light nuclei, the distribution is modified, namely to the 3-point Fermi distribution for oxygen and to the modified harmonic oscillator distribution for carbon~\cite{DEVRIES1987495}.
In case of oxygen, an additional model is used, approximating the nucleus as a tetrahedral cluster of $\alpha$-particles. This model leads to the geometry where the nuclear profile is not maximal at the centre, but its maximum is shifted to the outer shell of the nucleus. 

In addition to the aforementioned changes with respect to the proton solution in Ref.~\cite{Cepila:2025rkn}, the upper bound of the dipole-amplitude sampling range is extended to \mbox{$r_{\rm max} = 10^4~\mathrm{GeV}^{-1}$}, $b_{\rm max} = 10^4~\mathrm{GeV}^{-1}$, as can be seen in Fig.~\ref{fig:N_p_vs_A}. This is to address that in the interaction with large nuclei, a significant contribution comes from larger dipoles than in the case of protons. The remaining numerical setup for the BK equation solution is the same as in Ref.~\cite{Cepila:2025rkn}. 

\subsection{Tetrahedral oxygen}
The simulation of the oxygen nucleus as a cluster of $\alpha$-particles was performed by combining phenomenological models of its internal tetrahedral composition and a parametrisation of the helium density, obtained from elastic electron scattering experiments~\cite{DEVRIES1987495, Behera:2021zhi}. 

\begin{figure}
    \centering
    \includegraphics[width=.85\linewidth]{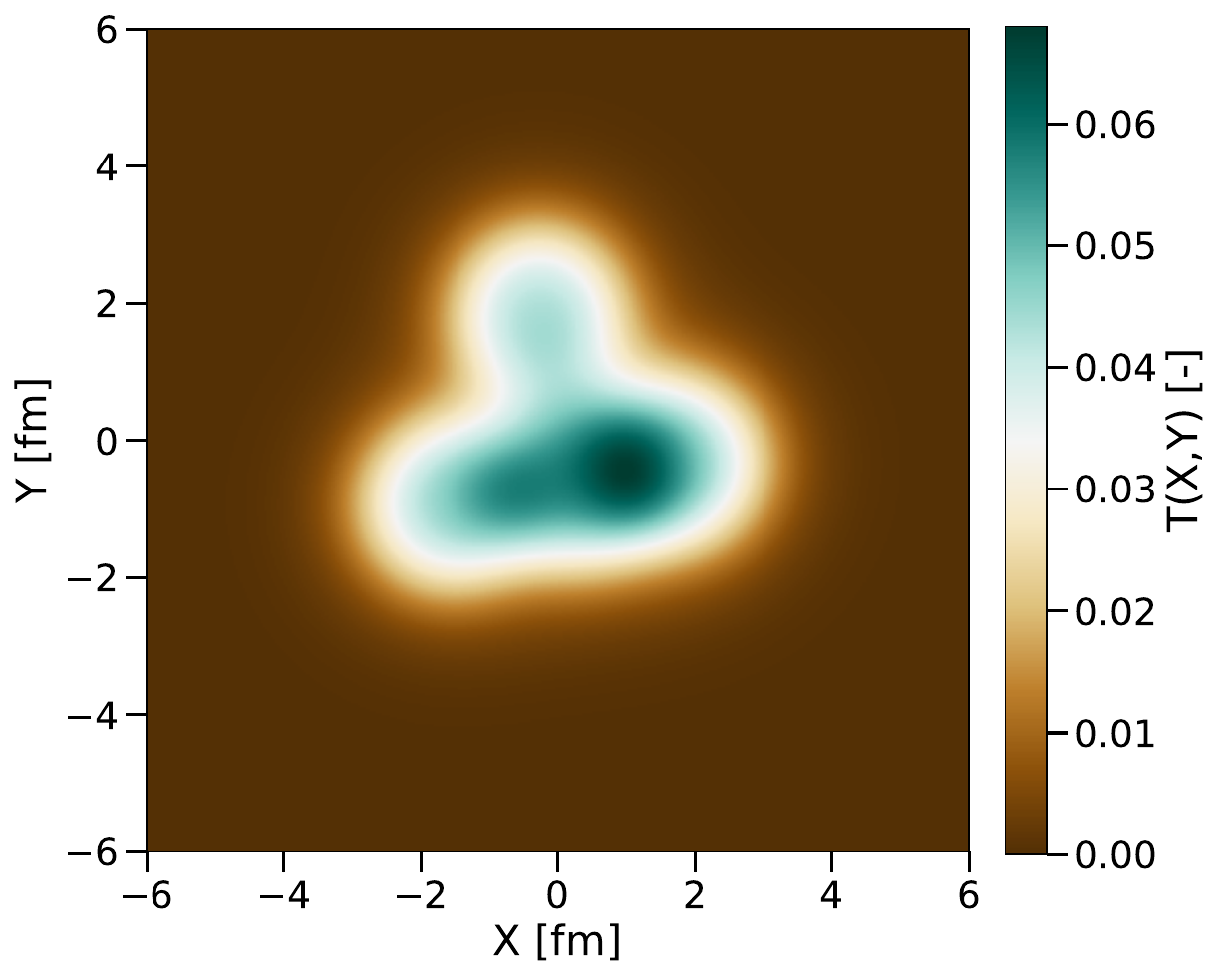}
    \caption{\label{fig:alpha-oxygen-density} 
    An example density profile obtained from a single random orientation of the oxygen $\alpha$-cluster tetrahedron.
    }
\end{figure}

We started by randomly orienting a tetrahedron with a side of 3.42\,fm that represents the $\alpha$-cluster forming the oxygen nucleus. This tetrahedron size has been shown to reproduce the $^{16}$O rms radius of 2.699\,fm~\cite{Li:2020vrg} as discussed in more detail in Ref.~\cite{Behera:2021zhi}. 

We imposed a stochastic uncertainty on the exact structure of the tetrahedron by applying a normal smearing on the position of the vertices of the tetrahedron in 3D with $\sigma$ = 0.1\,fm in order not to assume this structure to be absolutely rigid event by event.

\begin{figure}[!ht]
    \centering
    \includegraphics[width=.75\linewidth]{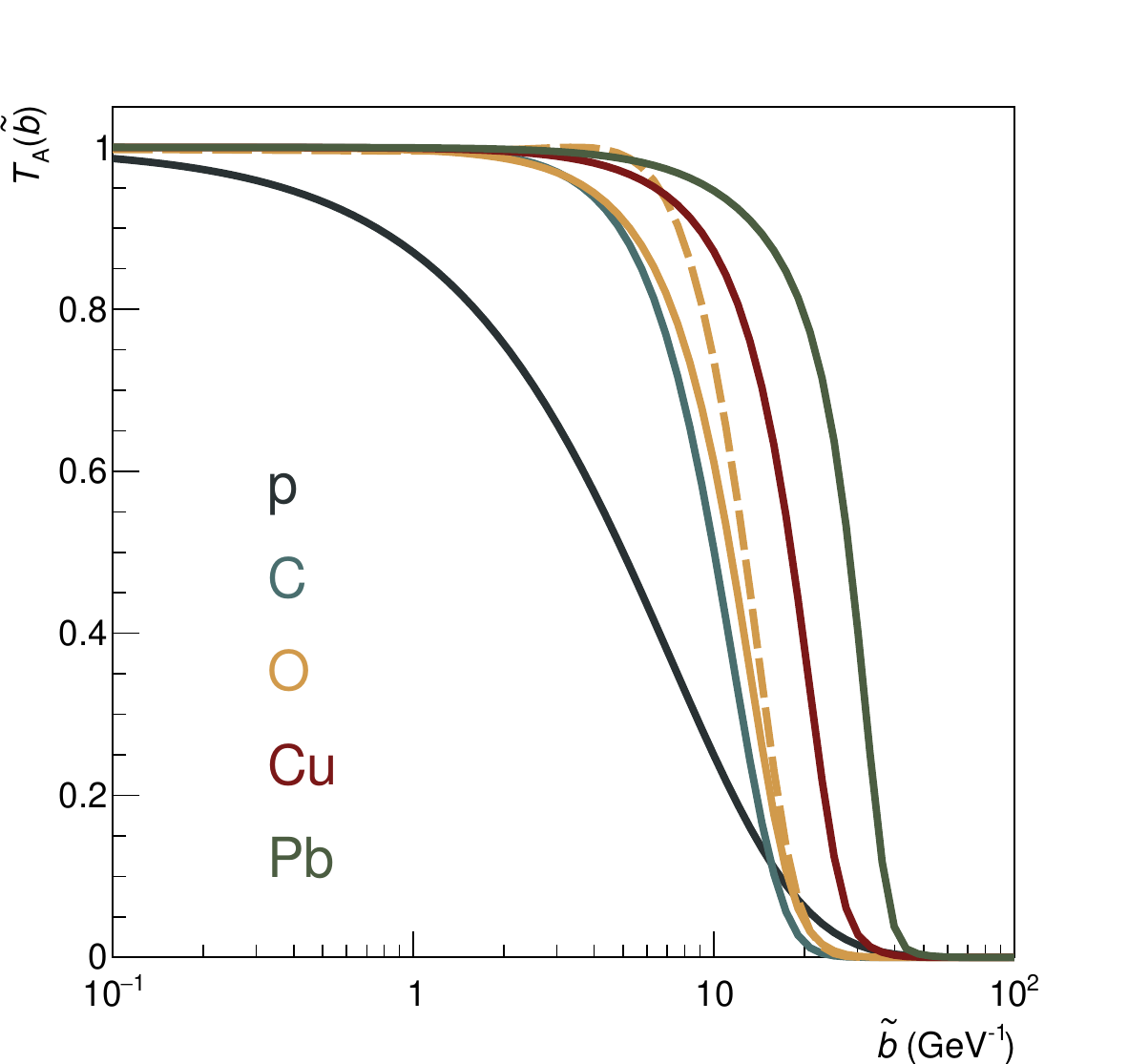}
    \caption{\label{fig:thickness_functions} 
    The Gaussian profile of the proton (shown in black), and partially integrated Woods-Saxon profiles of carbon, copper and lead (teal, red and green, respectively). Two versions of the oxygen profile are shown. The full ochre curve is obtained using the Woods-Saxon distribution, and the dashed ochre curve corresponds to the averaged tetrahedral oxygen formed by clusters of $\alpha$ particles.
    }
\end{figure}

Having generated the spatial tetrahedron structure, we projected its vertices onto the transverse plane, obtaining the centres of the helium nuclei in 2D. Their density was then modeled with the experimentally obtained parameterisation available in Table\,V of Ref.~\cite{DEVRIES1987495}. We integrated this distribution over the longitudinal direction, solving for the transverse density profile of the oxygen nucleus; an example shown in Fig.~\ref{fig:alpha-oxygen-density}.

In order to obtain a thickness function of the $\alpha$-clustered oxygen, we took an average over 5000 independent random setups of such a nucleus. The resulting thickness functions are shown in Fig.~\ref{fig:thickness_functions}.

\section{BK equation solutions\label{sec:solutions}}
\begin{figure}[!ht]
    \centering
    \includegraphics[width=.49\linewidth]{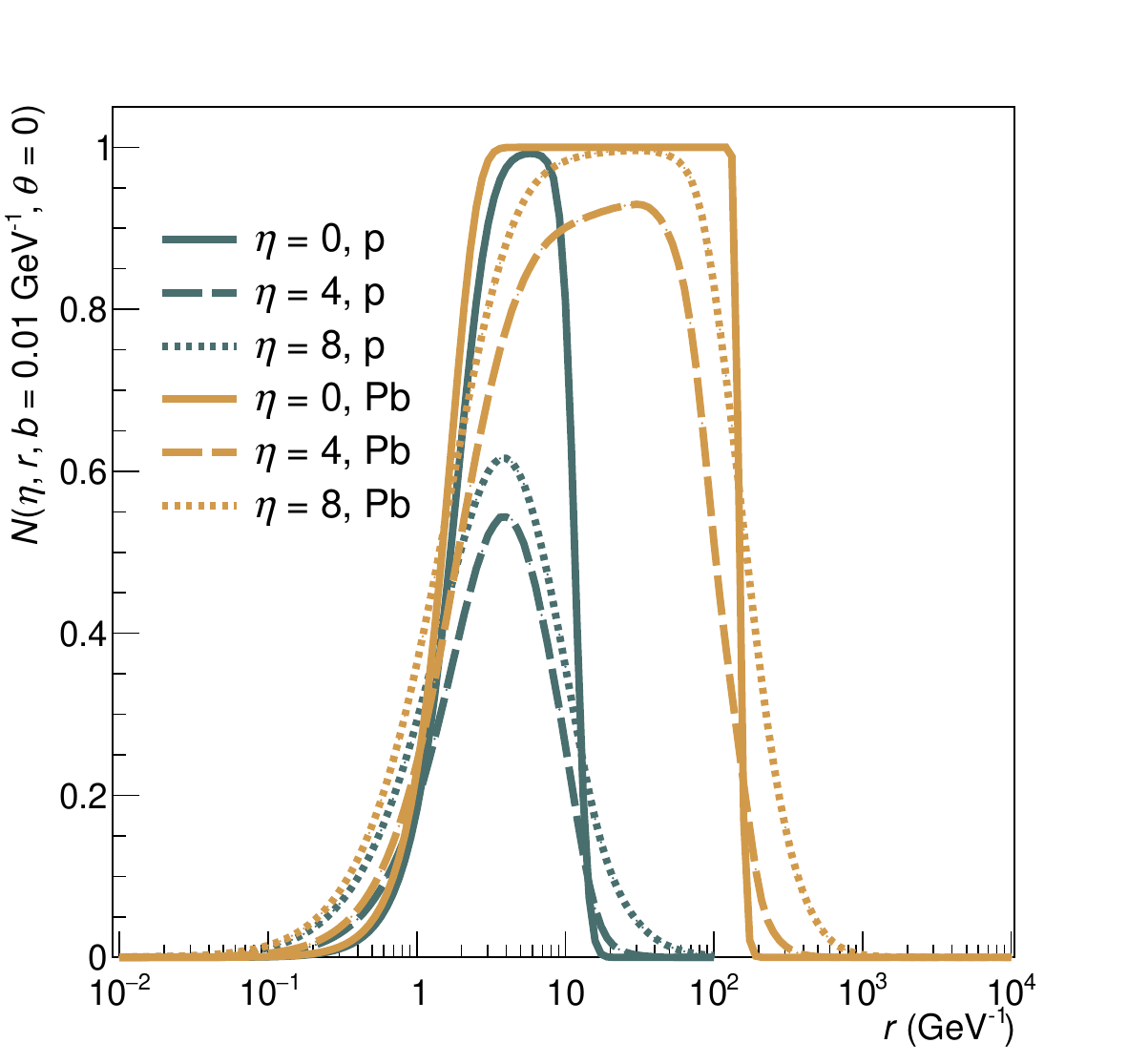}
    \includegraphics[width=.49\linewidth]{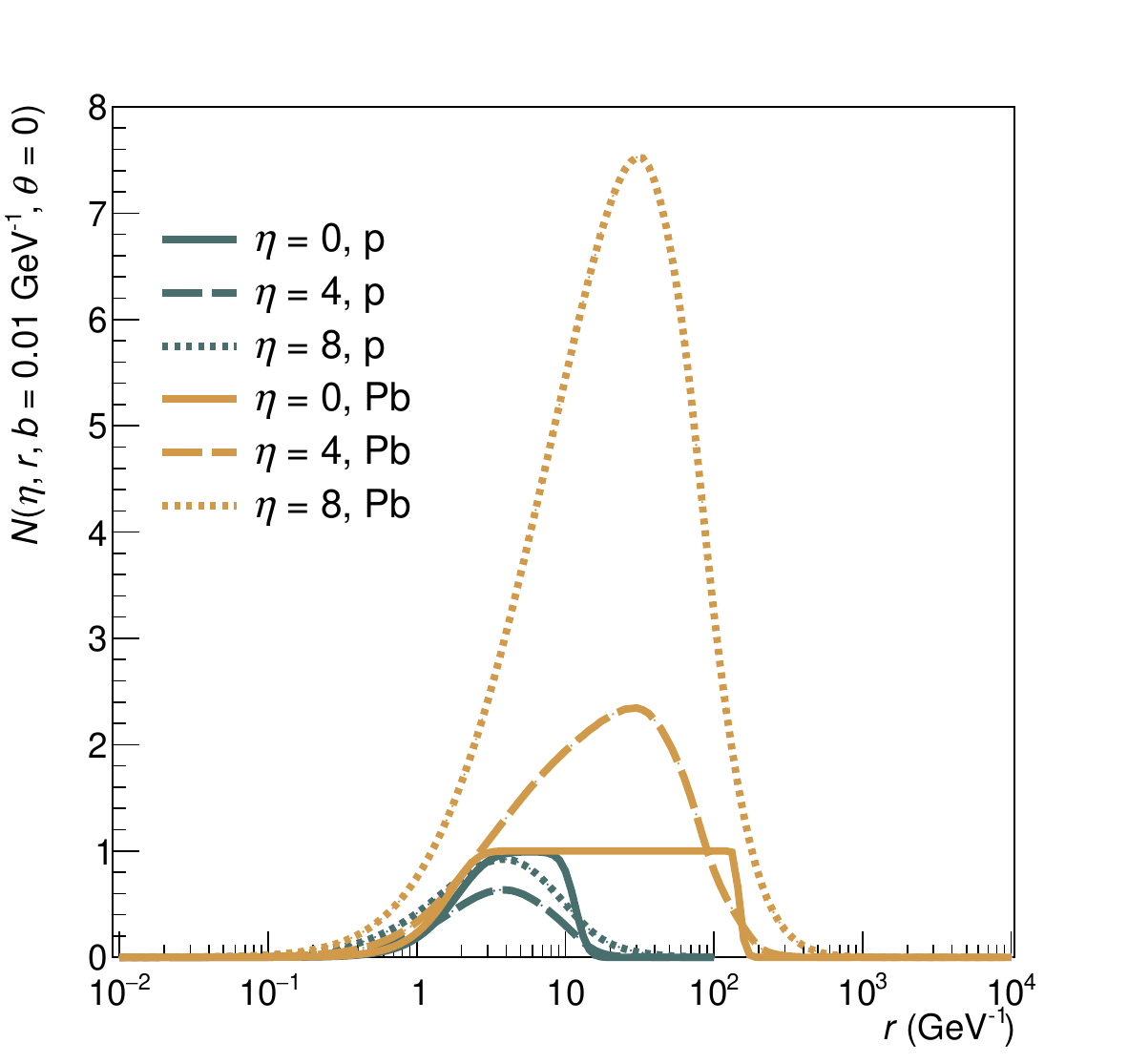}
    \caption{\label{fig:N_p_vs_A} Comparison of the dipole amplitudes at the initial condition (full) and evolved to rapidity 4 (dashed) and 8 (dotted) for a proton (teal) and lead (ochre) target. The shown amplitudes are central (i.e. at small impact parameter $b = 0.01~\mathrm{GeV}^{-1}$) and at $\theta = 0$ for the standard BK equation $\kappa = 1$ (left panel) and its linearised version $\kappa = 0$ (right panel).
    }
\end{figure}

The solution of the BK equation for the proton and nuclear targets is shown in Fig.~\ref{fig:N_p_vs_A}. For the case of the non-linear version of the equation with $\kappa = 1$, the evolution does not inflate the amplitude indefinitely, but rather modifies its morphology with a slower propagation speed of the wavefront. This alteration of the evolution speed was reported previously in the work focusing on the evolution in the magnitude of the vectors $\vec r$ and $\vec b$~\cite{Bendova:2019psy}. In contrast, the amplitude obtained by solving the linearised version of the BK equation grows steadily during the evolution. This behaviour has a significant impact on the observables calculated within both limiting cases. 

Note that while, in the particular phase-space point shown in Fig.~3, the peak value of the dipole amplitude decreases between the initial condition and $\eta = 4$, the physically relevant evolution takes place at smaller values of $r$, to the left of the peak. 
Values of $r \geq 1/\Lambda_{\text{QCD}}$ correspond to the non-perturbative region, and this part of phase-space is suppressed by the convolution with (mostly perturbatively calculated) wave-functions in calculations of measurable observables.
Here, the amplitude increases steadily, and the evolution resembles the standard travelling wave” known from solutions of the \mbox{only-$r$-dependent} BK equation. Furthermore, it is this low-$r$ region that has the dominant effect on the observables such as the structure functions, which consistently increase with rapidity, despite the decrease of the amplitude peak. This feature is natural to impact-parameter-dependent solutions of the BK equation and has been reported in Ref.~\cite{Bendova:2019psy}.

\section{Observables\label{sec:sf}}
To study the effect of saturation in the evolution of hadron structure, two processes are considered here: deep-inelastic scattering and coherent vector-meson photoproduction. 

\subsection{Nuclear structure functions}
For deep-inelastic scattering, the studied observable is the structure function  $F_2(x, Q^2)$. It is given by the cross-sections of the interaction between the virtual photon $\gamma^*$ and the target nucleus $A$ summed over possible photon polarisations (L, T) and flavours $f$ of the contributing quarks~\cite{Mueller:1989st, Nikolaev:1990ja}
\begin{align}
    F_2^{\rm p,\it A} (x, Q^2) &= \sum\limits_{f} \frac{Q^2}{4\pi^2 \aem} \left[ \sigma^{\gamma^* \rm p, \it A}_{\rm L, \it f}\left(Q^2, x_f\right) + \sigma^{\gamma^*\rm p,\it A}_{\rm T, \it f}\left(Q^2, x_f\right) \right],
    \label{eq:F2}
\end{align}
where $\aem$ is the electromagnetic coupling constant and 
\begin{align}
    x_f = \frac{x}{1+4\frac{m_f^2}{Q^2}} = \frac{x_0 e^{-\eta}}{1+4\frac{m_f^2}{Q^2}},
\end{align}
with $m_f$ the mass of a quark with flavour $f$. Virtuality of the exchanged photon is denoted as $Q^2$.

\begin{figure}[!ht]
    \centering
    \includegraphics[width=.49\linewidth]{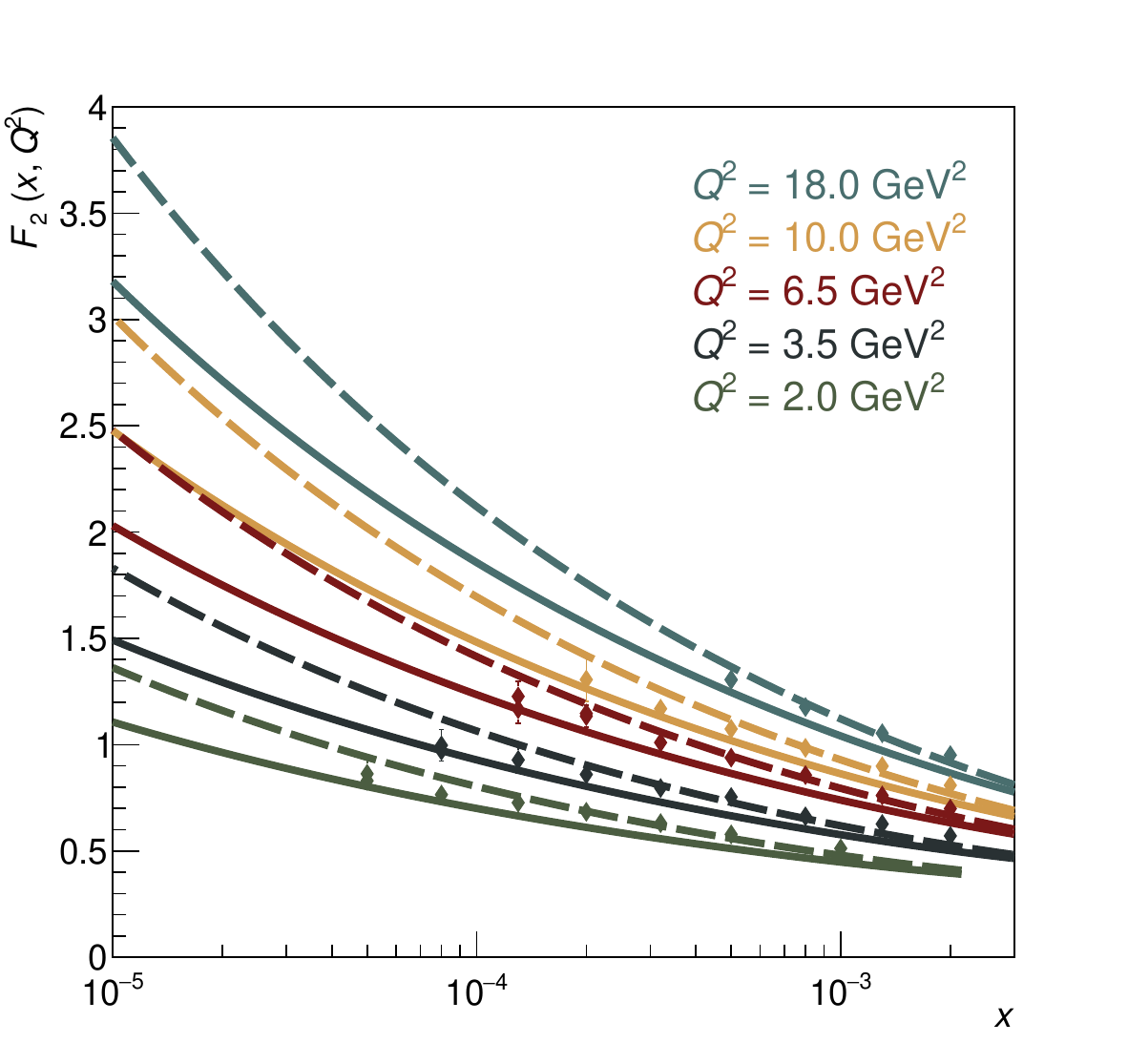}
    \includegraphics[width=.49\linewidth]{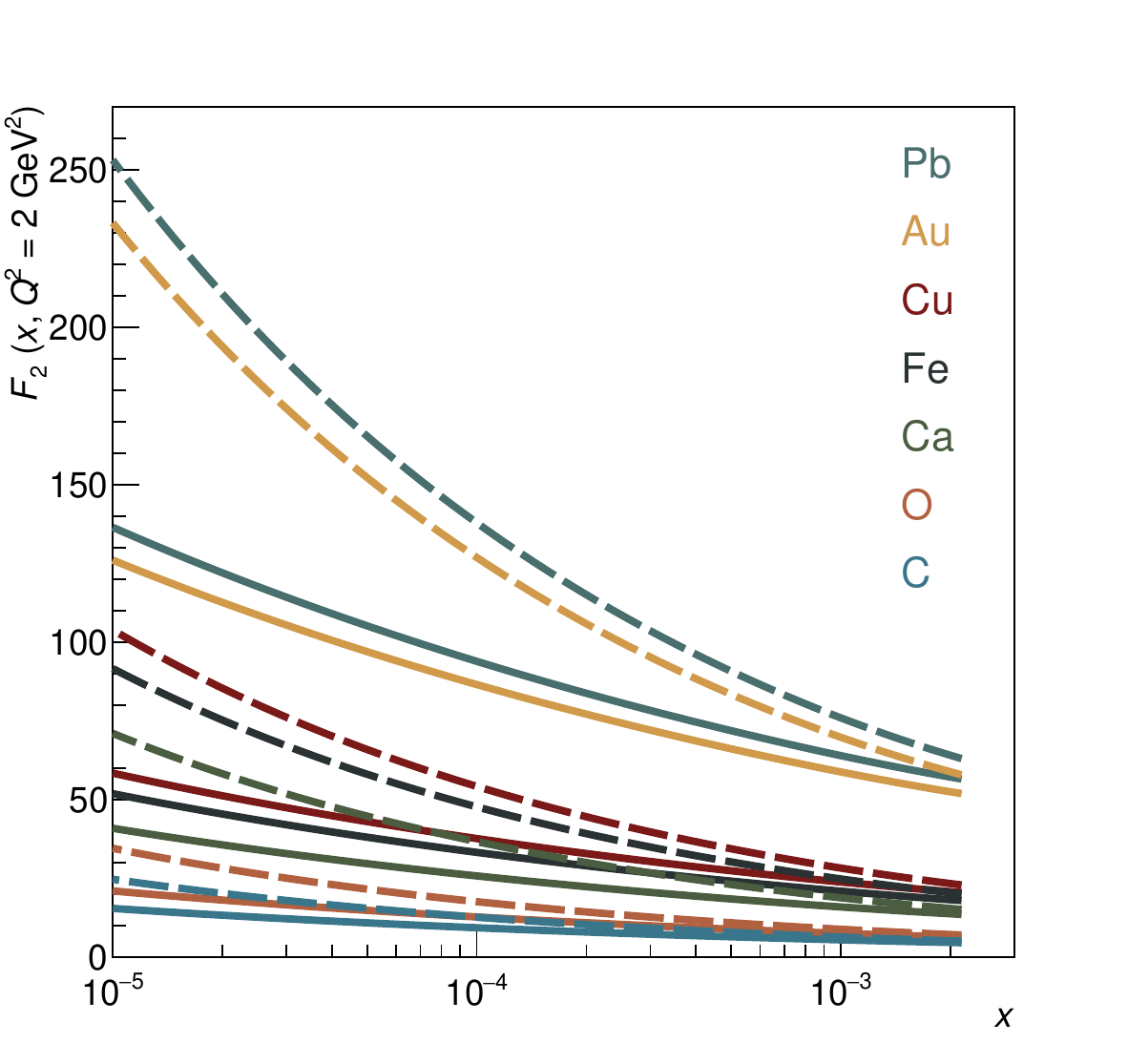}
    \caption{\label{fig:F2} 
    The $F_2 (x, Q^2)$ structure functions of proton (left panel) and nuclear (right panel) targets from the full-impact-parameter dependent BK equation. The available data from HERA~\cite{Aaron:2009aa} are shown for the proton target. Results for the tetrahedral oxygen are not shown in these plots as they are indistinguishable from the Woods-Saxon profile. The full curves correspond to the standard BK equation ($\kappa = 1$), the dashed ones come from the linearised version ($\kappa = 0$).
    }
\end{figure}

The photon--nucleus cross-sections are connected to the dipole amplitude by a convolution with the light-front photon wave functions as
\begin{align}
    \sigma^{\gamma^*\rm p,\it A}_{\rm L, T, \it f}(Q^2, x_f) = 2\pi\int\de{r} r \int\de{z} |\psi_{\rm L, T, \it f}(r, z, Q^2)|^2 \int \de{\vec b}~2N^{\rm p,\it A}(r, b, \theta, \eta),
\end{align}
with
\begin{align*}
    |\psi_{\rm L, \it f}(r, z, Q^2)|^2 &= \frac{3 \aem}{2\pi^2}e_f^2 4 Q^2 z^2 (1-z)^2 K_0^2\left(r \epsilon \right), \\
    |\psi_{\rm T, \it f}(r, z, Q^2)|^2 &= \frac{3 \aem}{2\pi^2}e_f^2 \left[(z^2 + (1-z)^2) \epsilon^2 K_1^2\left(r \epsilon \right) + m_f^2 K_0^2\left(r \epsilon \right) \right],
\end{align*}
where $K_{0, 1}(r\epsilon)$ are Bessel functions, $z$ is the fraction of photon momentum carried by one of the quarks from the dipole and $\epsilon = \sqrt{z(1-z)Q^2 + m_f^2}$.
Following Eq.~(\ref{eq:F2}), the structure functions of protons and a variety of nuclei were calculated and are shown in Fig.~\ref{fig:F2} for both limiting cases of the evolution equation, i.e. with $\kappa=0$ (dashed) and $\kappa=1$ (solid). 
The proton results for different scales are validated against the available experimental data from HERA~\cite{Aaron:2009aa}. One can see that both the non-linear and linearised BK model describe the available data on the proton target reasonably well, as measurements at much lower Bjorken-$x$ would be needed for their clear distinction. The nuclear results are shown in the right panel of Fig.~\ref{fig:F2}, ranging from light to heavy nuclei. Here, the difference between the non-linear and linearised model grows significantly with the mass number $A$. The reason is that in heavy nuclei, gluon clouds from different nucleons have a higher chance of overlapping and recombining, which in turn leads to stronger suppression from the non-linear component of the BK equation. 

\begin{figure}[!ht]
    \centering
    \includegraphics[width=.49\linewidth]{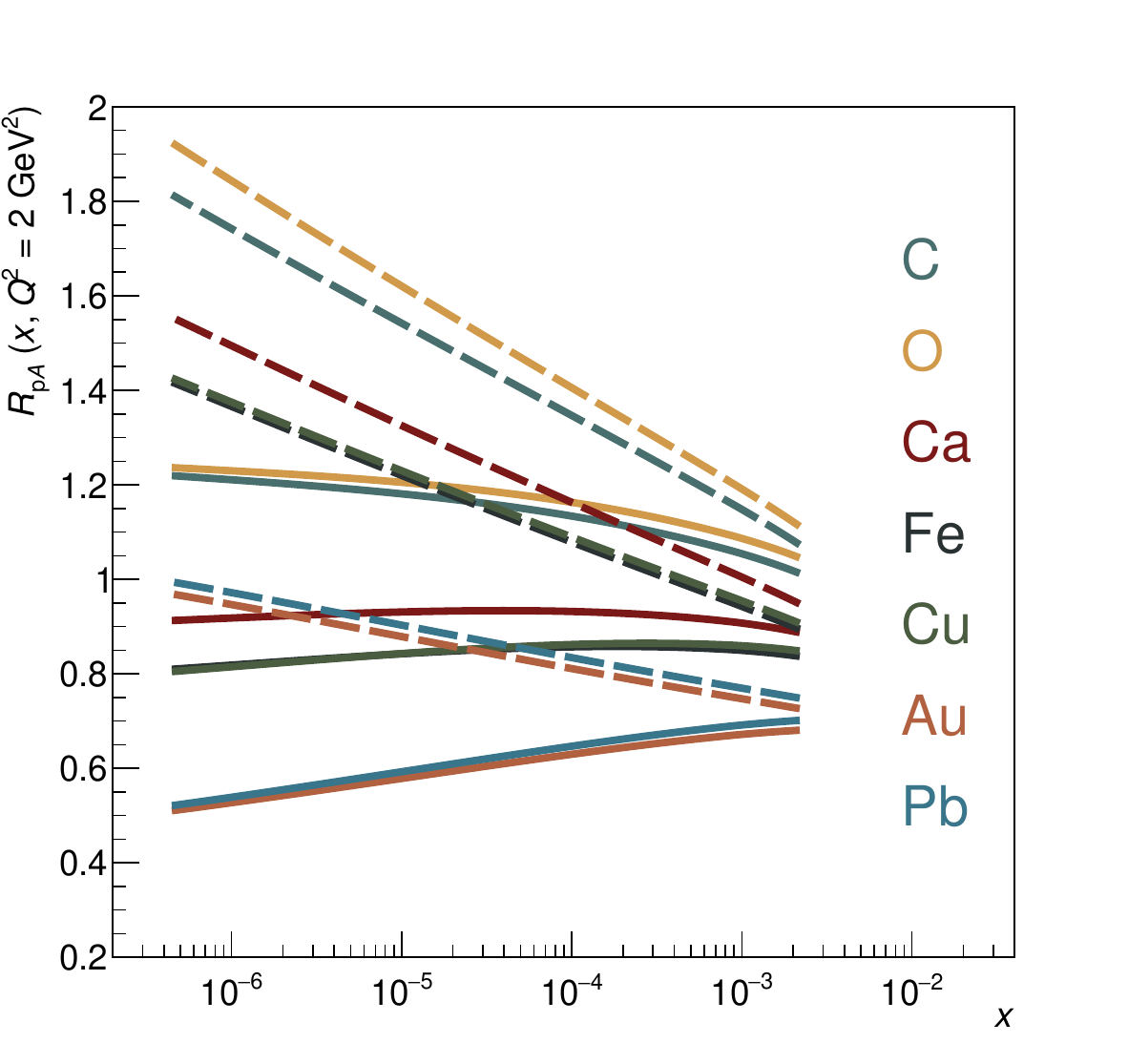}
    \includegraphics[width=.49\linewidth]{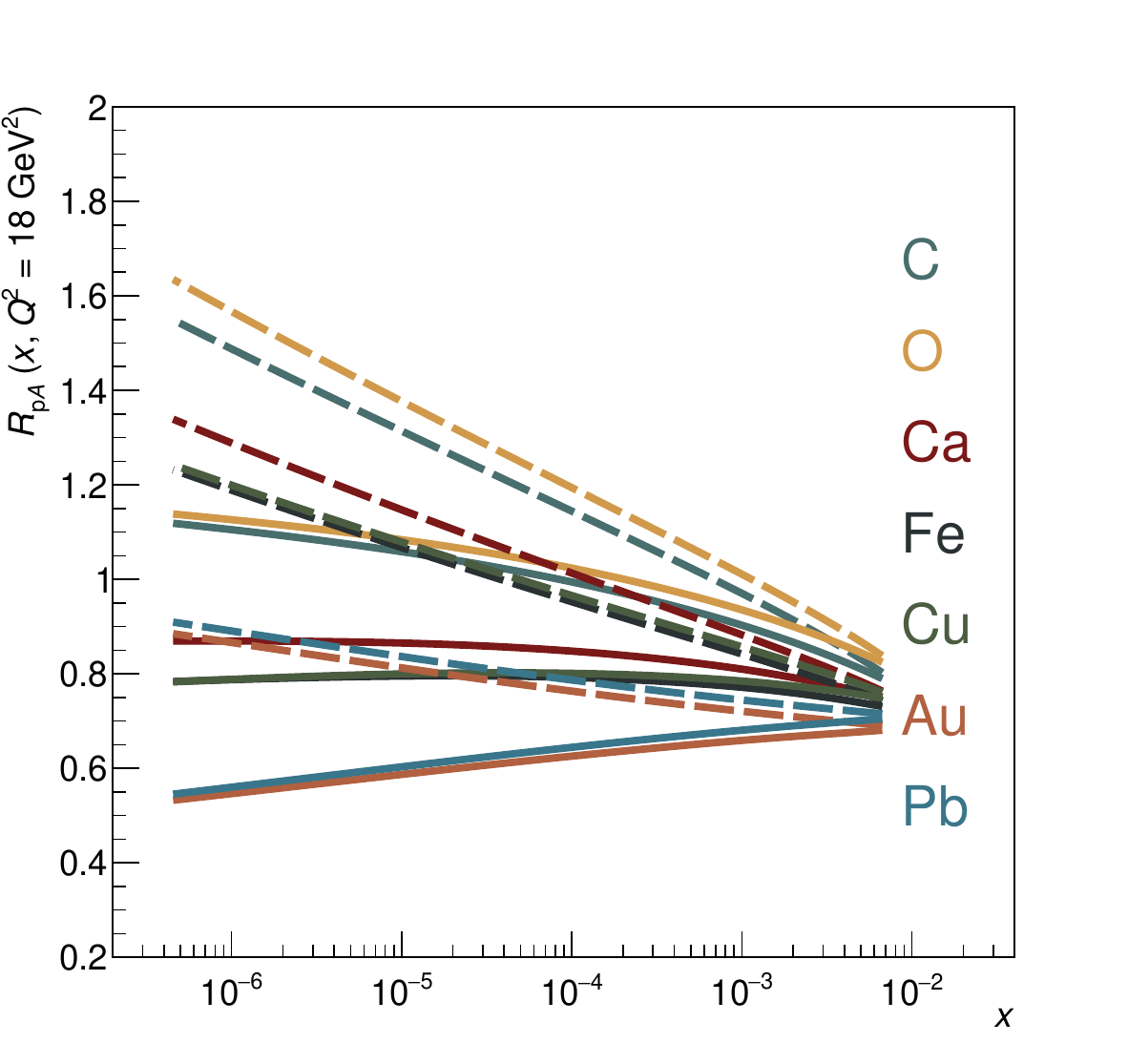}
    \caption{\label{fig:RpA} 
    The nuclear modification factor for virtualities $Q^2 = 2 \rm~GeV^{2}$ (left panel) and $Q^2 = 18~\rm GeV^{2}$ (right panel). The full curves correspond to the standard BK equation ($\kappa = 1$), while the dashed ones come from the linearised version ($\kappa = 0$).
    }
\end{figure}

In Fig.~\ref{fig:RpA}, predictions for the nuclear modification factor \mbox{$\RpA = F^A_{2} / A F_{2}$} are shown for two values of virtuality \mbox{$Q^2=2~\mathrm{GeV}^2$} (left panel) and \mbox{$Q^2=18~\mathrm{GeV}^2$} (right panel) for both the non-linear and linearised model. 
The nuclear modification factors obtained from the linearised version of BK (dashed) show no suppression and are enhanced for all nuclei in the whole range of Bjorken-$x$. For the non-linear version, the strongest suppression of the nuclear modification factor can be seen in heavy nuclei, confirming the importance of collective effects in densely packed systems. Note that the amount of suppression does not scale precisely with $A$ and e.g. $R_{\rm pO} > R_{\rm pC}$. This is because the thickness function enters the initial condition normalised such that $\max \big(T(\tilde b)\big) = 1$, emphasising its morphology rather than magnitude. The normalisation of the nuclear profile is then compensated by the explicit dependence of the saturation scale on $A$, leading to the correct low-energy limit of the nuclear modification factor with $\RpA \rightarrow 1$.

\subsection{Vector-meson production}
For coherent vector-meson production, both the total and the Mandelstam-$t$ differential cross section of $J/\psi$ photoproduction have been studied. In general, the differential cross-section of the diffractive exclusive vector meson production as a function of the Mandelstam $|t|$, photon virtuality $Q^2$, and the photon-ion centre-of-mass energy $W$ is given by the corresponding transverse and longitudinal amplitudes
\begin{align}
\frac{\de \sigma^{\rm p, \it A}}{\de |t|}(|t|, Q^2, W) =
\frac{1}{16\pi} \sum\limits_{\rm T, L} (1 + \beta^2_{\rm T, L}) R^2 _{\rm T, L} \left| \mathcal{A}^{\rm p, \it A}_{\rm T,L} \right|^2,
\end{align}
which are in turn connected to the dipole scattering amplitude using the optical theorem as 
\begin{align}
\mathcal{A}^{\rm p, \it A} = i \int \de r~ 2\pi r\int\limits_{0}^{1} \frac{\de z}{4\pi} \int \de \vec{b} \left( \Psi_V^\dagger \Psi\right)_{\rm T, L}e^{-i[\vec{b}- (\frac{1}{2}-z)\vec{r}]\vec{\Delta}} 2 N^{\rm p, \it A}\left(r,b,\theta, \eta\right). 
\end{align}
The magnitude of $\vec\Delta$ is given by $\vec\Delta^2=|t|$ and $\left( \Psi_V^\dagger \Psi\right)_{\rm T, L}$ is the overlap between the wave functions of a virtual photon and the vector meson~$V$ 
\begin{eqnarray}
\left(\Psi^{\dagger}_{V}\Psi\right)_{\rm T} &=& e_f\frac{\NC}{\pi z(1-z)}\bigg(m_f^{2} K_0(\epsilon r) \Phi_{\rm T}(r,z) \nonumber \\
&&-(z^{2}+(1-z)^{2})\epsilon K_1(\epsilon r)\partial_r\Phi_{\rm T}(r,z)\bigg)\nonumber\\
\left(\Psi^{\dagger}_{\rm V}\Psi\right)_{\rm L}&=&e_f\frac{\NC}{\pi}2Qz(1-z)K_0(\epsilon r)\nonumber\\
&&\times\left(m_V\Phi_{\rm L}(r,z)+\delta\frac{m_f^{2}-\nabla_r^{2}}{m_Vz(1-z)}\Phi_{\rm L}(r,z)\right),    
\end{eqnarray}
see e.g. Refs.~\cite{Kowalski:2006hc, Cox:2009ag}.
Here $ e_f$ is an effective charge that corresponds to the choice of the vector meson, $ \nabla^{2}_r=\frac{1}{r}\partial_r+\partial^{2}_r $ and $ \delta $ is a switch that enables including the non-local part of the wave function introduced in Refs.~\cite{Nemchik:1996cw, Nemchik:1994fp}. The scalar part of the vector meson wave function $\Phi_{\rm T, L}(r,z)$ is obtained using the boosted Gaussian model~\cite{Nemchik:1996cw, Nemchik:1994fp} with the same parameter setup as in Ref.~\cite{Cepila:2025rkn}. The corrections for the real part of the amplitude~$\beta^2_{\rm T, L}$~\cite{Forshaw:2003ki} and skewedness effects~$R^2 _{\rm T, L}$~\cite{Armesto:2014sma} are calculated in the same way as in~\cite{Cepila:2025rkn}. The rapidity relates to Bjorken-$x$ as 
 \begin{align}
     \eta = \ln \left( \frac{x_0}{x} \right)= \ln \left( x_0 \frac{W^2 + Q^2}{m_V^2 + Q^2} \right),
 \end{align}
 where $m_V$ is the mass of the emergent vector meson.

\begin{figure}[!ht]
    \centering
    \includegraphics[width=.45\linewidth]{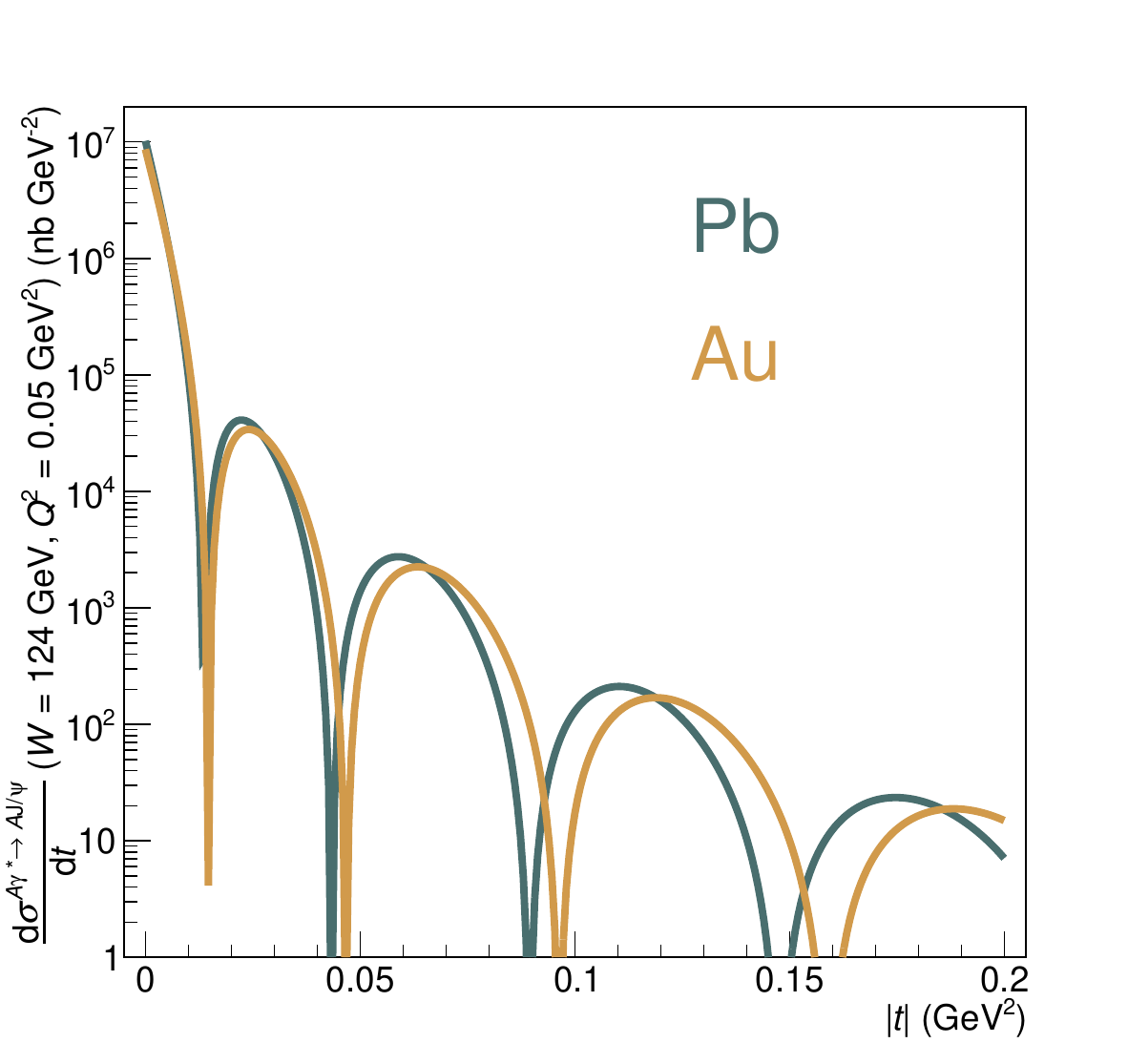}
    \includegraphics[width=.45\linewidth]{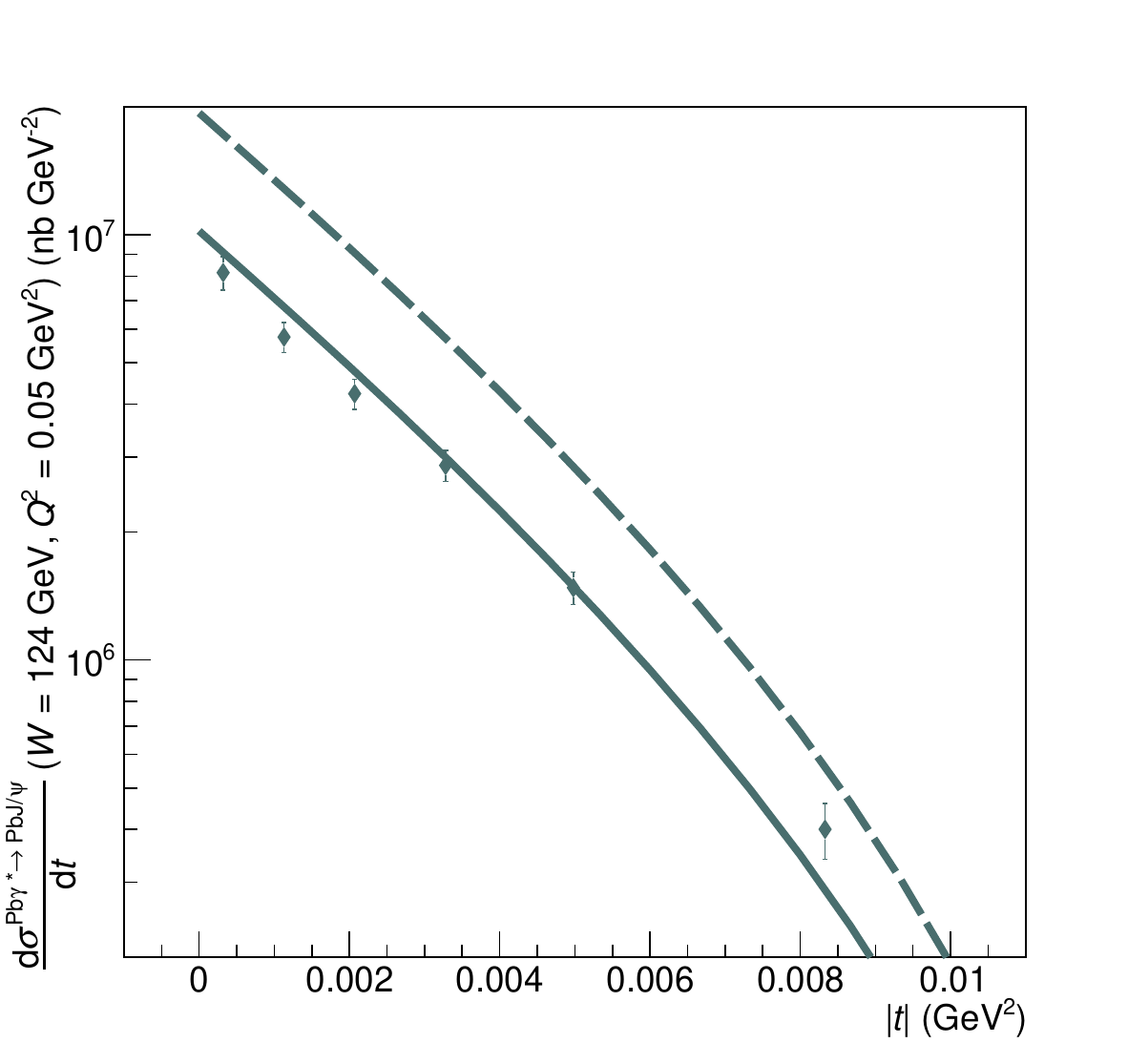}
    \caption{\label{fig:Jpsi_diff_h}
    Differential cross-sections of diffractive coherent $J/\psi$ photoproduction as a function of Madelstamm $t$ for heavy nuclei. Predictions for available data from ALICE~\cite{ALICE:2021tyx} are shown for the lead target in the right panel with the full curve. The dashed curve corresponds to the linearised version of the BK equation with $\kappa = 0$.
    }
\end{figure}

The available solutions of the BK equation have been used to obtain predictions for the coherent $J/\Psi$ photoproduction differential cross-sections for $W=124\mathrm{~GeV}$ as shown in the left panel of Fig.~\ref{fig:Jpsi_diff_h} for heavy nuclei. The expected series of dips is present, with the dip positions slightly changing based on a particular nucleus. 

In the right panel of Fig.~\ref{fig:Jpsi_diff_h}, one can see that the model with included saturation (non-linear BK) correctly describes available data from ALICE~\cite{ALICE:2021tyx}, and can also be used for other nuclei. The linearised version of the BK equation shows a sizeable discrepancy with the data. Compared to the DIS case, the logarithmic difference is, however, constant with $|t|$, suggesting that this dependence is not sensitive to the non-linear/linearised scenario at fixed energy. Note that no additional free parameter has been added to the model when moving from the proton target to the nuclear one. 

\begin{figure}[!ht]
    \centering
    \includegraphics[width=.45\linewidth]{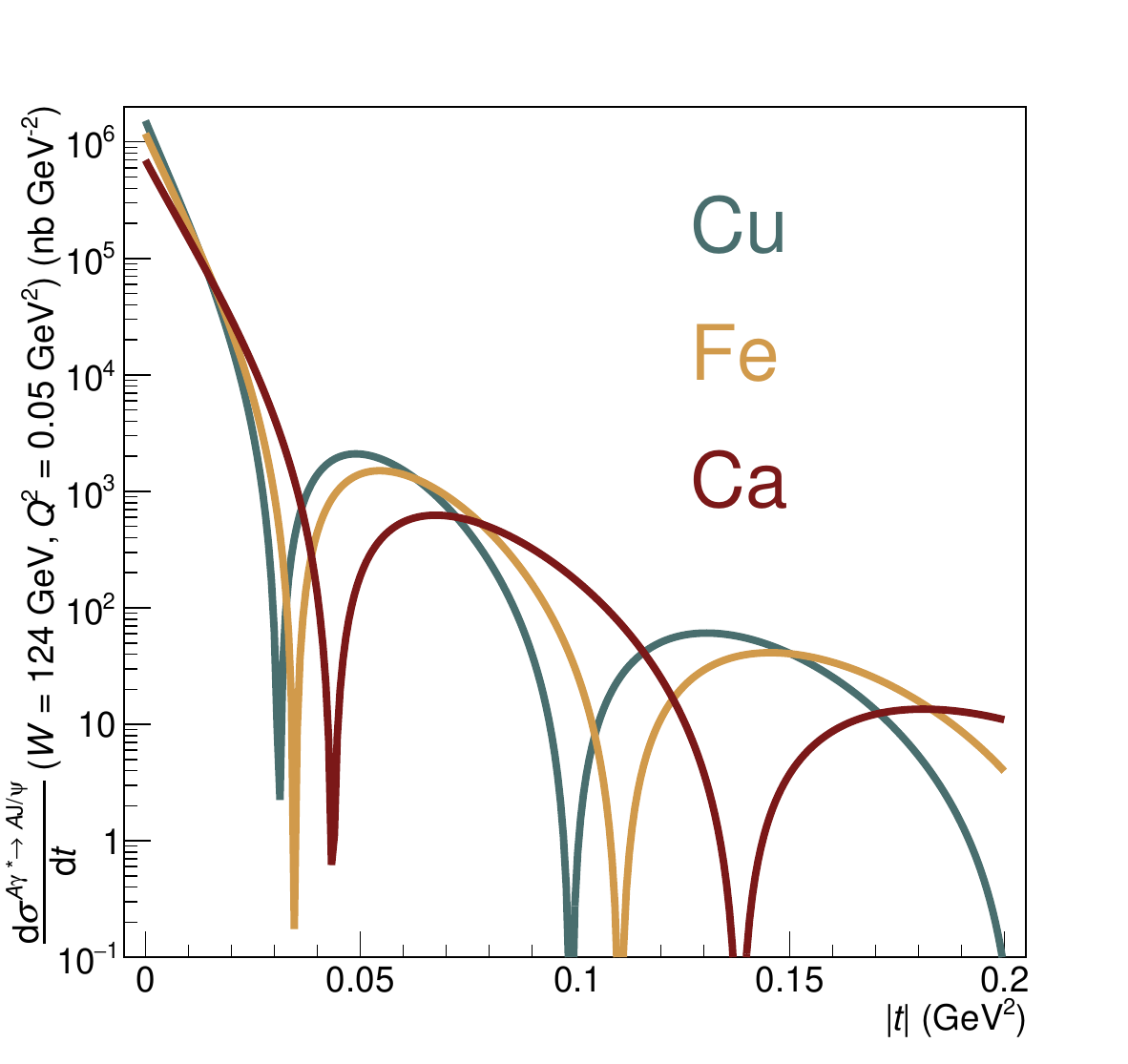}
    \includegraphics[width=.45\linewidth]{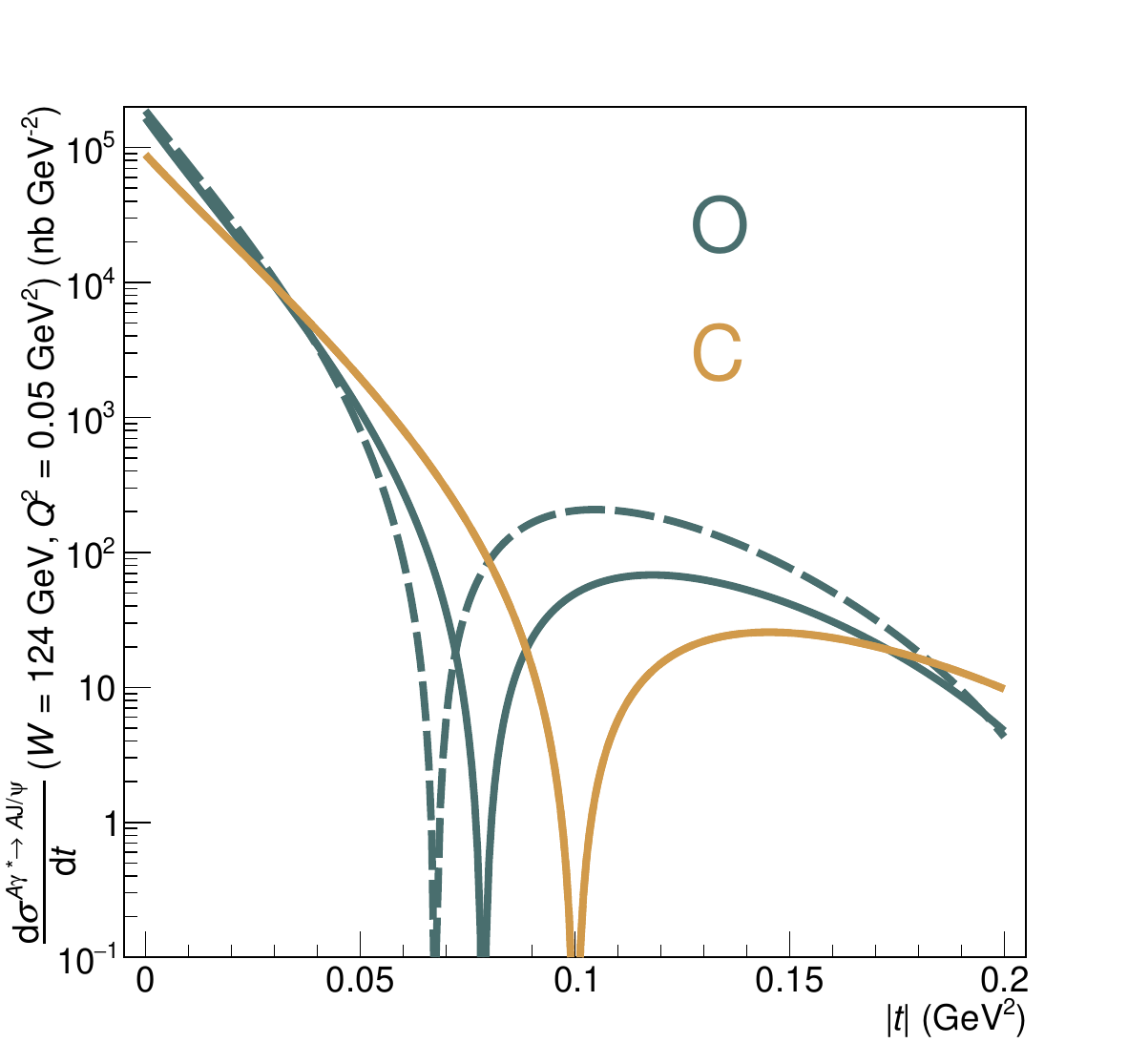}
    \caption{\label{fig:Jpsi_diff}
    Differential cross-sections of the diffractive $J/\psi$ photoproduction as a function of Madelstamm $t$ for mid-mass (left) and light (right) nuclei. The tetrahedral oxygen target is shown with the dashed curve.
    }
\end{figure}

The predictions for the $J/\Psi$ differential photoproduction cross-sections are shown in Fig.~\ref{fig:Jpsi_diff} for mid-mass nuclei (left panel) and light nuclei (right panel). The location of the diffractive dips reflects the different sizes of target nuclei, where the larger the nucleus, the earlier in $|t|$ the dips appear. The positions also move more significantly for lighter nuclei than we have seen for heavy nuclei. In the case of oxygen, the difference between the 3-point Fermi model (solid) and the tetrahedral helium model (dashed) is sizeable at large $|t|$, while at the level of the first dip, the models are not different enough to be distinguished by the foreseen experiments.   

\begin{figure}[!ht]
    \centering
    \includegraphics[width=.49\linewidth]
    {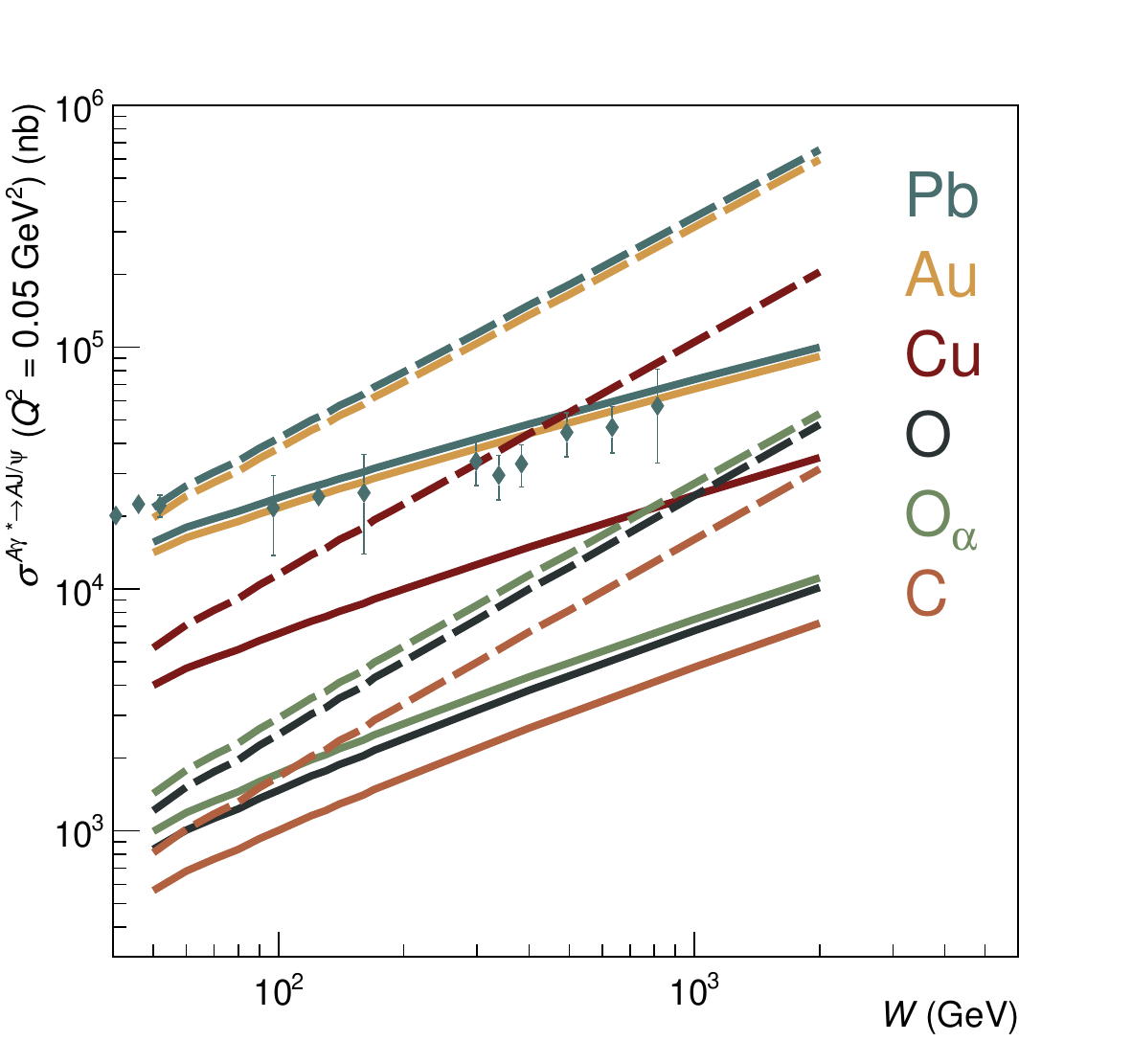}
    \includegraphics[width=.49\linewidth]
    {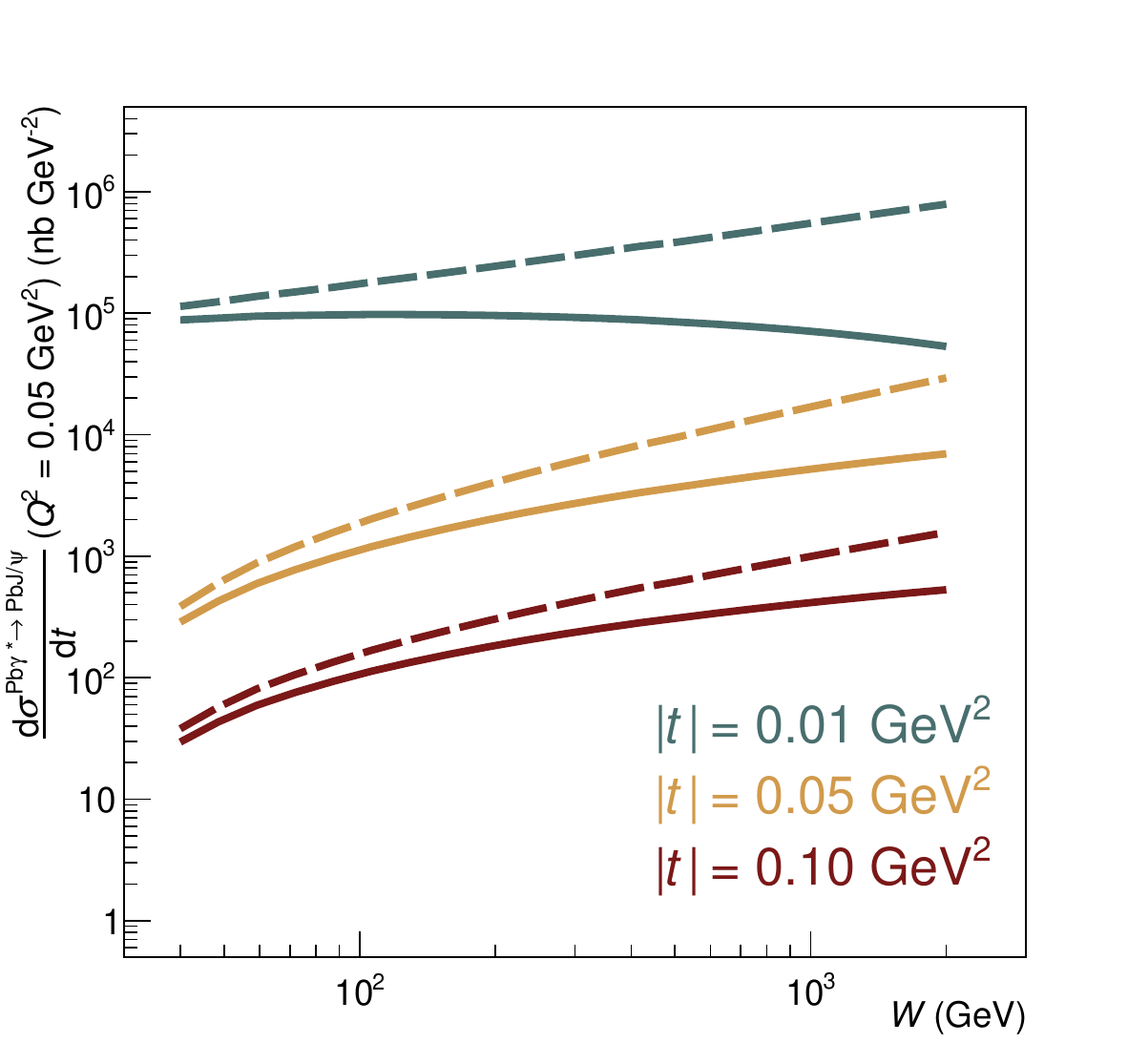}
    \caption{\label{fig:Jpsi_tot}
    Total (left) and differential (right) cross sections for $J/\psi$ photoproduction as functions of the $\gamma$–$A$ centre-of-mass energy $W$. The differential cross sections are shown for three fixed bins in $|t|$. The total cross sections are integrated over $|t| \in [0, 0.2]$ and are presented for various nuclei, together with available measurements from ALICE~\cite{ALICE:2023jgu} and CMS~\cite{CMS:2023snh} (both for lead). Solid curves correspond to the standard BK equation ($\kappa = 1$), while dashed curves represent the linearised case ($\kappa = 0$).
    }
\end{figure}

The predictions for the $J/\Psi$ total photoproduction cross-sections as a function of $W$, the energy of the $\gamma$-$A$ system, are shown in the left panel of Fig.~\ref{fig:Jpsi_tot} for various nuclei. The total cross-section is calculated as an integral of the differential one over Mandelstam-$|t|$ over the interval $[0, 0.2]$. The Pb predictions are compared to measurements from ALICE~\cite{ALICE:2023jgu}, showing a good agreement of the model with the available data. The dashed curves correspond to the linearised version of BK with $\kappa=0$. In contrast to the DIS case, the difference between non-linear/linearised approaches is significant even at low energies, and in order to describe available data, the non-linear suppression is important. 

The predictions for the cross-sections of differential $J/\Psi$ photoproduction off Pb are shown as a function of $W$ in the right panel of Fig.~\ref{fig:Jpsi_tot} for three specific bins in $|t|$. The clear difference between the non-linear and linearised model can be seen at $|t| \rightarrow 0$, where the non-linear model predicts a drop in the cross-section, while the linearised one only results in growth regardless of $|t|$. At the same time, the total cross-section for the non-linear model in the left panel of Fig.~\ref{fig:Jpsi_tot} rises with energy, suggesting that the fall-off of the contribution at $|t| \rightarrow 0$ is compensated by the rise of the contribution at larger values of~$|t|$. 
This behaviour demonstrates the crucial role of multi-differential measurements in elucidating the origin of saturation effects.

\section{Summary\label{sec:summary}}
In this paper, we have extended the solutions of the Balitsky-Kovchegov equation with the dependence on the full impact parameter, recently published in~\cite{Cepila:2025rkn}, from proton to nuclear targets. This was done solely by modifying the initial condition of the evolution, without presenting any additional free parameters. The dipole scattering amplitude includes nuclear effects based on the gluon component of the hadronic structure, and the behaviour of these effects comes purely from the dynamics of the BK equation. Here, the main studied phenomenon is the gluon saturation manifesting itself at low \mbox{Bjorken-$x$}. The resulting amplitude has been used to calculate observables from \mbox{deep-inelastic} scattering (structure function $F_2$) and from vector-meson production (cross-sections of coherent $J/\Psi$ photoproduction), all for a variety of nuclei. The observables have been confronted with available data, showing a good agreement. Predictions for different nuclei have been presented, being of potential interest for future experimental facilities such as EIC and LHeC, as well as for the current LHC program. 

Moreover, a simplified version of the BK equation without the non-linear term has been studied. This linearised version resembles the BFKL evolution equation, and even though the agreement with the available data for proton targets is reasonable in both cases, the difference is a lot more apparent for nuclear targets, allowing for the finding of an experimentally accessible observable that allows the discrimination between the models. The linearised version of the BK equation predicts a too steep rise of the cross-sections with energy (or equivalently with decreasing Bjorken-$x$). This rise does not allow for a reasonable prediction of the nuclear modification factor in the case of DIS and a correct data description in the case of the coherent $J/\Psi$ photoproduction cross-sections. Potentially, the most prominent signature showing an altered behaviour of the non-linear and linearised BK model is the $W$-distribution of $J/\Psi$ at different \mbox{Mandelstam-$t$}. For the linearised version, the cross-section rises with energy at all values of $|t|$ while the non-linear BK evolution predicts a drop in the cross-section at small $|t|$.

Finally, the effect of the alternative density profile for oxygen, composed of a tetrahedral combination of four helium nuclei, has been studied and compared to the more standard 3-point Fermi distribution. While the alternative model predicts a peak of the transverse density shifted more to the outer part of the nucleus, the effect on the studied observables is not pronounced. The only measurable difference between the two models is for the differential $J/\Psi$ photoproduction cross-sections at large $|t|$, where each of the models predicts a different position of the dips. The code to generate the tetrahedral $\alpha$ clusters for the oxygen is publicly available for further use. 

\section*{Oxygen modeling software availability}
The nuclear helper software to generate nuclear profiles is publicly available at~\cite{nuclear_helper}.

\section*{Acknowledgements}
This work was partially funded by the Czech Science Foundation (GAČR), Project No. 22-27262S.
This work was supported by the Ministry of Education, Youth and Sports of the Czech Republic through the e-INFRA CZ (ID:90254).
M. Matas acknowledges the support by the CTU Mobility Project MSCA-F-CZ-III (Reg. No. CZ.02.01.01/00/22\_010/0008601).
\bibliography{bibliography}

@article{Forshaw:2003ki,
    author = "Forshaw, Jeffrey R. and Sandapen, R. and Shaw, Graham",
    title = "{Color dipoles and rho, phi electroproduction}",
    eprint = "hep-ph/0312172",
    archivePrefix = "arXiv",
    doi = "10.1103/PhysRevD.69.094013",
    journal = "Phys. Rev. D",
    volume = "69",
    pages = "094013",
    year = "2004"
}

@article{Armesto:2014sma,
    author = "Armesto, N\'estor and Rezaeian, Amir H.",
    title = "{Exclusive vector meson production at high energies and gluon saturation}",
    eprint = "1402.4831",
    archivePrefix = "arXiv",
    primaryClass = "hep-ph",
    doi = "10.1103/PhysRevD.90.054003",
    journal = "Phys. Rev. D",
    volume = "90",
    number = "5",
    pages = "054003",
    year = "2014"
}

@article{Cox:2009ag,
    author = "Cox, B. E. and Forshaw, J. R. and Sandapen, R.",
    title = "{Diffractive upsilon production at the LHC}",
    eprint = "0905.0102",
    archivePrefix = "arXiv",
    primaryClass = "hep-ph",
    reportNumber = "MAN-HEP-2008-45",
    doi = "10.1088/1126-6708/2009/06/034",
    journal = "JHEP",
    volume = "06",
    pages = "034",
    year = "2009"
}

@article{Nemchik:1994fp,
    author = "Nemchik, J. and Nikolaev, Nikolai N. and Zakharov, B. G.",
    title = "{Scanning the BFKL pomeron in elastic production of vector mesons at HERA}",
    eprint = "hep-ph/9405355",
    archivePrefix = "arXiv",
    reportNumber = "KFA-IKP-TH-1994-17",
    doi = "10.1016/0370-2693(94)90314-X",
    journal = "Phys. Lett. B",
    volume = "341",
    pages = "228--237",
    year = "1994"
}

@article{Nemchik:1996cw,
    author = "Nemchik, J. and Nikolaev, Nikolai N. and Predazzi, E. and Zakharov, B. G.",
    title = "{Color dipole phenomenology of diffractive electroproduction of light vector mesons at HERA}",
    eprint = "hep-ph/9605231",
    archivePrefix = "arXiv",
    reportNumber = "DFTT-71-95, KFA-IKP-TH-95-24",
    doi = "10.1007/s002880050448",
    journal = "Z. Phys. C",
    volume = "75",
    pages = "71--87",
    year = "1997"
}

@article{Kowalski:2006hc,
    author = "Kowalski, H. and Motyka, L. and Watt, G.",
    title = "{Exclusive diffractive processes at HERA within the dipole picture}",
    eprint = "hep-ph/0606272",
    archivePrefix = "arXiv",
    reportNumber = "DESY-06-095",
    doi = "10.1103/PhysRevD.74.074016",
    journal = "Phys. Rev. D",
    volume = "74",
    pages = "074016",
    year = "2006"
}

@article{Nikolaev:1990ja,
    author = "Nikolaev, Nikolai N. and Zakharov, B. G.",
    editor = "Khalatnikov, I. M. and Mineev, V. P.",
    title = "{Color transparency and scaling properties of nuclear shadowing in deep inelastic scattering}",
    reportNumber = "OUTP-90-23-P",
    doi = "10.1007/BF01483577",
    journal = "Z. Phys. C",
    volume = "49",
    pages = "607--618",
    year = "1991"
}

@article{Mueller:1989st,
    author = "Mueller, Alfred H.",
    title = "{Small x Behavior and Parton Saturation: A QCD Model}",
    reportNumber = "CU-TP-441a",
    doi = "10.1016/0550-3213(90)90173-B",
    journal = "Nucl. Phys. B",
    volume = "335",
    pages = "115--137",
    year = "1990"
}

@article{Bendova:2022xhw,
    author = "Bendova, D. and Cepila, J. and Gon\c{c}alves, V. P. and Sena, C. R.",
    title = "{Deeply virtual Compton scattering at the EIC and LHeC: a comparison among saturation approaches}",
    doi = "10.1140/epjc/s10052-022-10059-9",
    journal = "Eur. Phys. J. C",
    volume = "82",
    number = "2",
    pages = "99",
    year = "2022"
}

@article{DEVRIES1987495,
title = {Nuclear charge-density-distribution parameters from elastic electron scattering},
journal = {Atomic Data and Nuclear Data Tables},
volume = {36},
number = {3},
pages = {495-536},
year = {1987},
issn = {0092-640X},
doi = {https://doi.org/10.1016/0092-640X(87)90013-1},
url = {https://www.sciencedirect.com/science/article/pii/0092640X87900131},
author = {H. {De Vries} and C.W. {De Jager} and C. {De Vries}},
}

@article{Cepila:2020xol,
    author = "Cepila, J. and Contreras, J. G. and Matas, M.",
    title = "{Predictions for nuclear structure functions from the impact-parameter dependent Balitsky-Kovchegov equation}",
    eprint = "2002.11056",
    archivePrefix = "arXiv",
    primaryClass = "hep-ph",
    doi = "10.1103/PhysRevC.102.044318",
    journal = "Phys. Rev. C",
    volume = "102",
    number = "4",
    pages = "044318",
    year = "2020"
}

@article{Berger:2011ew,
    author = "Berger, Jeffrey and Stasto, Anna M.",
    title = "{Small x nonlinear evolution with impact parameter and the structure function data}",
    eprint = "1106.5740",
    archivePrefix = "arXiv",
    primaryClass = "hep-ph",
    doi = "10.1103/PhysRevD.84.094022",
    journal = "Phys. Rev. D",
    volume = "84",
    pages = "094022",
    year = "2011"
}

@article{Albacete:2010sy,
    author = "Albacete, Javier L. and Armesto, Nestor and Milhano, Jose Guilherme and Quiroga-Arias, Paloma and Salgado, Carlos A.",
    title = "{AAMQS: A non-linear QCD analysis of new HERA data at small-x including heavy quarks}",
    eprint = "1012.4408",
    archivePrefix = "arXiv",
    primaryClass = "hep-ph",
    doi = "10.1140/epjc/s10052-011-1705-3",
    journal = "Eur. Phys. J. C",
    volume = "71",
    pages = "1705",
    year = "2011"
}

@article{Gribov:1984tu,
      author         = "Gribov, L. V. and Levin, E. M. and Ryskin, M. G.",
      title          = "{Semihard Processes in QCD}",
      journal        = "Phys. Rept.",
      volume         = "100",
      year           = "1983",
      pages          = "1-150",
      doi            = "10.1016/0370-1573(83)90022-4",
      SLACcitation   = "%%CITATION = PRPLC,100,1;%%"
}

@article{Kovchegov:1999yj,
    author = "Kovchegov, Yuri V.",
    title = "{Small x F(2) structure function of a nucleus including multiple pomeron exchanges}",
    eprint = "hep-ph/9901281",
    archivePrefix = "arXiv",
    reportNumber = "NUC-MN-99-1-T, TPI-MINN-99-05",
    doi = "10.1103/PhysRevD.60.034008",
    journal = "Phys. Rev. D",
    volume = "60",
    pages = "034008",
    year = "1999"
}

@article{Balitsky:1995ub,
    author = "Balitsky, I.",
    title = "{Operator expansion for high-energy scattering}",
    eprint = "hep-ph/9509348",
    archivePrefix = "arXiv",
    reportNumber = "MIT-CTP-2470",
    doi = "10.1016/0550-3213(95)00638-9",
    journal = "Nucl. Phys. B",
    volume = "463",
    pages = "99--160",
    year = "1996"
}

@article{Mueller:1985wy,
    author = "Mueller, Alfred H. and Qiu, Jian-wei",
    title = "{Gluon Recombination and Shadowing at Small Values of x}",
    reportNumber = "CU-TP-322",
    doi = "10.1016/0550-3213(86)90164-1",
    journal = "Nucl. Phys. B",
    volume = "268",
    pages = "427--452",
    year = "1986"
}

@article{Bendova:2019psy,
    author = "Bendova, D. and Cepila, J. and Contreras, J. G. and Matas, M.",
    title = "{Solution to the Balitsky-Kovchegov equation with the collinearly improved kernel including impact-parameter dependence}",
    eprint = "1907.12123",
    archivePrefix = "arXiv",
    primaryClass = "hep-ph",
    doi = "10.1103/PhysRevD.100.054015",
    journal = "Phys. Rev. D",
    volume = "100",
    number = "5",
    pages = "054015",
    year = "2019"
}

@article{Cepila:2018faq,
    author = "Cepila, J. and Contreras, J. G. and Matas, M.",
    title = "{Collinearly improved kernel suppresses Coulomb tails in the impact-parameter dependent Balitsky-Kovchegov evolution}",
    eprint = "1812.02548",
    archivePrefix = "arXiv",
    primaryClass = "hep-ph",
    doi = "10.1103/PhysRevD.99.051502",
    journal = "Phys. Rev. D",
    volume = "99",
    number = "5",
    pages = "051502",
    year = "2019"
}

@article{Lipatov:1976zz,
    author = "Lipatov, L. N.",
    title = "{Reggeization of the Vector Meson and the Vacuum Singularity in Nonabelian Gauge Theories}",
    journal = "Sov. J. Nucl. Phys.",
    volume = "23",
    pages = "338--345",
    year = "1976"
}

@article{Mueller:1993rr,
    author = "Mueller, Alfred H.",
    title = "{Soft gluons in the infinite momentum wave function and the BFKL pomeron}",
    reportNumber = "SLAC-PUB-10047, CU-TP-609",
    doi = "10.1016/0550-3213(94)90116-3",
    journal = "Nucl. Phys. B",
    volume = "415",
    pages = "373--385",
    year = "1994"
}

@article{Cepila:2023pvh,
    author = "Cepila, J. and Contreras, J. G. and Vaculciak, M.",
    title = "{Solutions to the Balitsky-Kovchegov equation including the dipole orientation}",
    eprint = "2309.02910",
    archivePrefix = "arXiv",
    primaryClass = "hep-ph",
    doi = "10.1016/j.physletb.2023.138360",
    journal = "Phys. Lett. B",
    volume = "848",
    pages = "138360",
    year = "2024"
}

@article{Cepila:2025rkn,
    author = "Cepila, J. and Contreras, J. G. and Vaculciak, M.",
    title = "{Exclusive quarkonium photoproduction: Predictions with the Balitsky-Kovchegov equation including the full impact-parameter dependence}",
    eprint = "2501.09462",
    archivePrefix = "arXiv",
    primaryClass = "hep-ph",
    doi = "10.1103/PhysRevD.111.056002",
    journal = "Phys. Rev. D",
    volume = "111",
    number = "5",
    pages = "056002",
    year = "2025"
}

@article{Achenbach:2023pba,
    author = "Achenbach, P. and others",
    title = "{The present and future of QCD}",
    eprint = "2303.02579",
    archivePrefix = "arXiv",
    primaryClass = "hep-ph",
    reportNumber = "JLAB-PHY-23-3808",
    doi = "10.1016/j.nuclphysa.2024.122874",
    journal = "Nucl. Phys. A",
    volume = "1047",
    pages = "122874",
    year = "2024"
}

@article{Kovchegov_2000,
   title={Unitarization of the BFKL Pomeron on a nucleus},
   volume={61},
   ISSN={1089-4918},
   url={http://dx.doi.org/10.1103/PhysRevD.61.074018},
   DOI={10.1103/physrevd.61.074018},
   number={7},
   journal={Physical Review D},
   publisher={American Physical Society (APS)},
   author={Kovchegov, Yuri V.},
   year={2000},
   month={3}
}

@article{Kovchegov_2007,
   title={Triumvirate of running couplings in small-x evolution},
   volume={784},
   ISSN={0375-9474},
   url={http://dx.doi.org/10.1016/j.nuclphysa.2006.10.075},
   DOI={10.1016/j.nuclphysa.2006.10.075},
   number={1-4},
   journal={Nuclear Physics A},
   publisher={Elsevier BV},
   author={Kovchegov, Yuri V. and Weigert, Heribert},
   year={2007},
   month={3},
   pages={188–226}
}

@article{Balitsky_2007,
   title={Quark contribution to the small-x evolution of color dipole},
   volume={75},
   ISSN={1550-2368},
   url={http://dx.doi.org/10.1103/PhysRevD.75.014001},
   DOI={10.1103/physrevd.75.014001},
   number={1},
   journal={Physical Review D},
   publisher={American Physical Society (APS)},
   author={Balitsky, Ian},
   year={2007},
   month={1}
}

@article{Ducloue:2019ezk,
    author = "Duclou\'e, B. and Iancu, E. and Mueller, A. H. and Soyez, G. and Triantafyllopoulos, D. N.",
    title = "{Non-linear evolution in QCD at high-energy beyond leading order}",
    eprint = "1902.06637",
    archivePrefix = "arXiv",
    primaryClass = "hep-ph",
    reportNumber = "INT-PUB-19-006",
    doi = "10.1007/JHEP04(2019)081",
    journal = "JHEP",
    volume = "04",
    pages = "081",
    year = "2019"
}

@article{ALICE:2023jgu,
    author = "Acharya, Shreyasi and others",
    collaboration = "ALICE",
    title = "{Energy dependence of coherent photonuclear production of J/\ensuremath{\psi} mesons in ultra-peripheral Pb-Pb collisions at $ \sqrt{{\textrm{s}}_{\textrm{NN}}} $ = 5.02 TeV}",
    eprint = "2305.19060",
    archivePrefix = "arXiv",
    primaryClass = "nucl-ex",
    reportNumber = "CERN-EP-2023-100",
    doi = "10.1007/JHEP10(2023)119",
    journal = "JHEP",
    volume = "10",
    pages = "119",
    year = "2023"
}

@article{Ryskin:1992ui,
    author = "Ryskin, M. G.",
    title = "{Diffractive J / psi electroproduction in LLA QCD}",
    reportNumber = "LU-TP-92-12",
    doi = "10.1007/BF01555742",
    journal = "Z. Phys. C",
    volume = "57",
    pages = "89--92",
    year = "1993"
}

@article{Brodsky:1994kf,
    author = "Brodsky, Stanley J. and Frankfurt, L. and Gunion, J. F. and Mueller, Alfred H. and Strikman, M.",
    title = "{Diffractive leptoproduction of vector mesons in QCD}",
    eprint = "hep-ph/9402283",
    archivePrefix = "arXiv",
    reportNumber = "SLAC-PUB-6412, CU-TP-617, UCD-93-36",
    doi = "10.1103/PhysRevD.50.3134",
    journal = "Phys. Rev. D",
    volume = "50",
    pages = "3134--3144",
    year = "1994"
}

@article{Bendova:2020hbb,
    author = "Bendova, D. and Cepila, J. and Contreras, J. G. and Matas, M.",
    title = "{Photonuclear $J/\psi$ production at the LHC: Proton-based versus nuclear dipole scattering amplitudes}",
    eprint = "2006.12980",
    archivePrefix = "arXiv",
    primaryClass = "hep-ph",
    doi = "10.1016/j.physletb.2021.136306",
    journal = "Phys. Lett. B",
    volume = "817",
    pages = "136306",
    year = "2021"
}

@article{Ivanov:2004ax,
    author = "Ivanov, I. P. and Nikolaev, N. N. and Savin, A. A.",
    title = "{Diffractive vector meson production at HERA: From soft to hard QCD}",
    eprint = "hep-ph/0501034",
    archivePrefix = "arXiv",
    reportNumber = "DESY-04-243",
    doi = "10.1134/S1063779606010011",
    journal = "Phys. Part. Nucl.",
    volume = "37",
    pages = "1--85",
    year = "2006"
}

@article{Newman:2013ada,
    author = "Newman, Paul and Wing, Matthew",
    title = "{The Hadronic Final State at HERA}",
    eprint = "1308.3368",
    archivePrefix = "arXiv",
    primaryClass = "hep-ex",
    doi = "10.1103/RevModPhys.86.1037",
    journal = "Rev. Mod. Phys.",
    volume = "86",
    number = "3",
    pages = "1037",
    year = "2014"
}

@article{Baltz:2007kq,
    author = "Baltz, A. J. and others",
    title = "{The Physics of Ultraperipheral Collisions at the LHC}",
    eprint = "0706.3356",
    archivePrefix = "arXiv",
    primaryClass = "nucl-ex",
    doi = "10.1016/j.physrep.2007.12.001",
    journal = "Phys. Rept.",
    volume = "458",
    pages = "1--171",
    year = "2008"
}

@article{Contreras:2015dqa,
    author = "Contreras, J. G. and Tapia Takaki, J. D.",
    title = "{Ultra-peripheral heavy-ion collisions at the LHC}",
    doi = "10.1142/S0217751X15420129",
    journal = "Int. J. Mod. Phys. A",
    volume = "30",
    pages = "1542012",
    year = "2015"
}

@article{Klein:2019qfb,
    author = {Klein, Spencer R. and M\"antysaari, Heikki},
    title = "{Imaging the nucleus with high-energy photons}",
    eprint = "1910.10858",
    archivePrefix = "arXiv",
    primaryClass = "hep-ex",
    doi = "10.1038/s42254-019-0107-6",
    journal = "Nature Rev. Phys.",
    volume = "1",
    number = "11",
    pages = "662--674",
    year = "2019"
}

@article{LHeCStudyGroup:2012zhm,
    author = "Abelleira Fernandez, J. L. and others",
    collaboration = "LHeC Study Group",
    title = "{A Large Hadron Electron Collider at CERN: Report on the Physics and Design Concepts for Machine and Detector}",
    eprint = "1206.2913",
    archivePrefix = "arXiv",
    primaryClass = "physics.acc-ph",
    reportNumber = "SLAC-R-999, CERN-OPEN-2012-015, LHEC-NOTE-2012-001-GEN",
    doi = "10.1088/0954-3899/39/7/075001",
    journal = "J. Phys. G",
    volume = "39",
    pages = "075001",
    year = "2012"
}

@article{Accardi:2012qut,
    author = "Accardi, A. and others",
    editor = "Deshpande, A. and Meziani, Z. E. and Qiu, J. W.",
    title = "{Electron Ion Collider: The Next QCD Frontier}: {Understanding the glue that binds us all}",
    eprint = "1212.1701",
    archivePrefix = "arXiv",
    primaryClass = "nucl-ex",
    reportNumber = "BNL-98815-2012-JA, JLAB-PHY-12-1652",
    doi = "10.1140/epja/i2016-16268-9",
    journal = "Eur. Phys. J. A",
    volume = "52",
    number = "9",
    pages = "268",
    year = "2016"
}

@article{AbdulKhalek:2021gbh,
    author = "Abdul Khalek, R. and others",
    title = "{Science Requirements and Detector Concepts for the Electron-Ion Collider}: {EIC Yellow Report}",
    eprint = "2103.05419",
    archivePrefix = "arXiv",
    primaryClass = "physics.ins-det",
    reportNumber = "BNL-220990-2021-FORE, JLAB-PHY-21-3198, LA-UR-21-20953",
    doi = "10.1016/j.nuclphysa.2022.122447",
    journal = "Nucl. Phys. A",
    volume = "1026",
    pages = "122447",
    year = "2022"
}

@article{Li:2020vrg,
    author = "Li, Yi-An and Zhang, Song and Ma, Yu-Gang",
    title = "{Signatures of $\alpha$-clustering in $^{16}$O by using a multiphase transport model}",
    eprint = "2010.10003",
    archivePrefix = "arXiv",
    primaryClass = "hep-ph",
    doi = "10.1103/PhysRevC.102.054907",
    journal = "Phys. Rev. C",
    volume = "102",
    number = "5",
    pages = "054907",
    year = "2020"
}

@article{Behera:2021zhi,
    author = "Behera, Debadatta and Mallick, Neelkamal and Tripathy, Sushanta and Prasad, Suraj and Mishra, Aditya Nath and Sahoo, Raghunath",
    title = "{Predictions on global properties in O+O collisions at the Large Hadron Collider using a multi-phase transport model}",
    eprint = "2110.04016",
    archivePrefix = "arXiv",
    primaryClass = "hep-ph",
    doi = "10.1140/epja/s10050-022-00823-6",
    journal = "Eur. Phys. J. A",
    volume = "58",
    number = "9",
    pages = "175",
    year = "2022"
}

@article{ALICE:2021tyx,
    author = "Acharya, Shreyasi and others",
    collaboration = "ALICE",
    title = "{First measurement of the |$t$|-dependence of coherent $J/\psi$ photonuclear production}",
    eprint = "2101.04623",
    archivePrefix = "arXiv",
    primaryClass = "nucl-ex",
    reportNumber = "CERN-EP-2021-003",
    doi = "10.1016/j.physletb.2021.136280",
    journal = "Phys. Lett. B",
    volume = "817",
    pages = "136280",
    year = "2021"
}

@article{Armesto:2006ph,
    author = "Armesto, Nestor",
    title = "{Nuclear shadowing}",
    eprint = "hep-ph/0604108",
    archivePrefix = "arXiv",
    doi = "10.1088/0954-3899/32/11/R01",
    journal = "J. Phys. G",
    volume = "32",
    pages = "R367--R394",
    year = "2006"
}

@article{Frankfurt:2011cs,
    author = "Frankfurt, L. and Guzey, V. and Strikman, M.",
    title = "{Leading Twist Nuclear Shadowing Phenomena in Hard Processes with Nuclei}",
    eprint = "1106.2091",
    archivePrefix = "arXiv",
    primaryClass = "hep-ph",
    reportNumber = "JLAB-THY-11-1379",
    doi = "10.1016/j.physrep.2011.12.002",
    journal = "Phys. Rept.",
    volume = "512",
    pages = "255--393",
    year = "2012"
}

@article{Kuraev:1977fs,
    author = "Kuraev, E. A. and Lipatov, L. N. and Fadin, Victor S.",
    title = "{The Pomeranchuk Singularity in Nonabelian Gauge Theories}",
    journal = "Sov. Phys. JETP",
    volume = "45",
    pages = "199--204",
    year = "1977"
}

@article{Balitsky:1978ic,
    author = "Balitsky, I. I. and Lipatov, L. N.",
    title = "{The Pomeranchuk Singularity in Quantum Chromodynamics}",
    journal = "Sov. J. Nucl. Phys.",
    volume = "28",
    pages = "822--829",
    year = "1978"
}

@article{CMS:2023snh,
    author = "Tumasyan, Armen and others",
    collaboration = "CMS",
    title = "{Probing Small Bjorken-x Nuclear Gluonic Structure via Coherent J/\ensuremath{\psi} Photoproduction in Ultraperipheral Pb-Pb Collisions at sNN=5.02\,\,TeV}",
    eprint = "2303.16984",
    archivePrefix = "arXiv",
    primaryClass = "nucl-ex",
    reportNumber = "CMS-HIN-22-002, CERN-EP-2023-031",
    doi = "10.1103/PhysRevLett.131.262301",
    journal = "Phys. Rev. Lett.",
    volume = "131",
    number = "26",
    pages = "262301",
    year = "2023"
}

@article{Mantysaari:2025ltq,
    author = {M{\"a}ntysaari, Heikki and Roch, Hendrik and Salazar, Farid and Schenke, Bj{\"o}rn and Shen, Chun and Zhao, Wenbin},
    title = "{Global Bayesian analysis of J/{\ensuremath{\psi}} photoproduction on proton and lead targets}",
    eprint = "2507.14087",
    archivePrefix = "arXiv",
    primaryClass = "hep-ph",
    doi = "10.1103/pcmz-dyz1",
    journal = "Phys. Rev. D",
    volume = "113",
    number = "1",
    pages = "014038",
    year = "2026"
}

@article{Goncalves:2025wwt,
    author = {Goncalves, Victor P. and Santana, Luana and Sch{\"a}fer, Wolfgang},
    title = "{Investigating the inclusive D0 photoproduction in ultraperipheral PbPb collisions at the Large Hadron Collider}",
    eprint = "2506.02223",
    archivePrefix = "arXiv",
    primaryClass = "hep-ph",
    doi = "10.1016/j.physletb.2025.139859",
    journal = "Phys. Lett. B",
    volume = "869",
    pages = "139859",
    year = "2025"
}

@misc{nuclear_helper,
  author = {Matas, M.},
  title = {Nuclear Helper},
  year = {2024},
  publisher = {GitHub},
  journal = {GitHub repository},
  howpublished = {\url{https://github.com/MatasMarek/nuclear_helper}},
  commit = {41354014320d3fae32c045e292bbca4f70fb036c}
}

@article{Aaij:2014iea,
	author         = "Aaij, Roel and others",
	title          = "{Updated measurements of exclusive J/Psi and Psi(2S)
	production cross-sections in pp collisions at $\sqrt{s}$ =
	7 TeV}",
	collaboration  = "LHCb collaboration",
	journal        = "J.Phys.",
	volume         = "G41",
	pages          = "055002",
	doi            = "10.1088/0954-3899/41/5/055002",
	year           = "2014",
	eprint         = "1401.3288",
	archivePrefix  = "arXiv",
	primaryClass   = "hep-ex",
	reportNumber   = "CERN-PH-EP-2013-233, LHCB-PAPER-2013-059",
	SLACcitation   = "%%CITATION = ARXIV:1401.3288;%%",
}

@article{Aaij:2015kea,
	author = "Aaij, Roel and others",
	collaboration = "LHCb",
	title = "{Measurement of the exclusive \ensuremath{\Upsilon} production cross-section in pp collisions at $ \sqrt{s}=7 $ TeV and 8 TeV}",
	eprint = "1505.08139",
	archivePrefix = "arXiv",
	primaryClass = "hep-ex",
	reportNumber = "LHCB-PAPER-2015-011, CERN-PH-EP-2015-123",
	doi = "10.1007/JHEP09(2015)084",
	journal = "JHEP",
	volume = "09",
	pages = "084",
	year = "2015"
}

@article{Aaron:2009aa,
	author         = "Aaron, F.D. and others",
	title          = "{Combined Measurement and QCD Analysis of the Inclusive
	e+- p Scattering Cross Sections at HERA}",
	collaboration  = "H1 and ZEUS Collaboration",
	journal        = "JHEP",
	volume         = "1001",
	pages          = "109",
	doi            = "10.1007/JHEP01(2010)109",
	year           = "2010",
	eprint         = "0911.0884",
	archivePrefix  = "arXiv",
	primaryClass   = "hep-ex",
	reportNumber   = "DESY-09-158",
	SLACcitation   = "%%CITATION = ARXIV:0911.0884;%%",
}

@article{Aaron:2009xp,
	author = "Aaron, F. D. and others",
	collaboration = "H1",
	title = "{Diffractive Electroproduction of rho and phi Mesons at HERA}",
	eprint = "0910.5831",
	archivePrefix = "arXiv",
	primaryClass = "hep-ex",
	reportNumber = "DESY-09-093",
	doi = "10.1007/JHEP05(2010)032",
	journal = "JHEP",
	volume = "05",
	pages = "032",
	year = "2010"
}

@article{Adloff:2000vm,
	author = "Adloff, C. and others",
	collaboration = "H1",
	title = "{Elastic photoproduction of J / psi and Upsilon mesons at HERA}",
	eprint = "hep-ex/0003020",
	archivePrefix = "arXiv",
	reportNumber = "DESY-00-037",
	doi = "10.1016/S0370-2693(00)00530-X",
	journal = "Phys. Lett. B",
	volume = "483",
	pages = "23--35",
	year = "2000"
}

@article{Alexa:2013xxa,
	author         = "Alexa, C. and others",
	title          = "{Elastic and Proton-Dissociative Photoproduction of J/psi Mesons at HERA}",
	collaboration  = "H1",
	journal        = "Eur. Phys. J.",
	volume         = "C73",
	year           = "2013",
	number         = "6",
	pages          = "2466",
	doi            = "10.1140/epjc/s10052-013-2466-y",
	eprint         = "1304.5162",
	archivePrefix  = "arXiv",
	primaryClass   = "hep-ex",
	reportNumber   = "DESY-13-058",
	SLACcitation   = "%%CITATION = ARXIV:1304.5162;%%"
}

@article{ALICE:2014eof,
    author = "Abelev, Betty Bezverkhny and others",
    collaboration = "ALICE",
    title = "{Exclusive $\mathrm{J/}\psi$ photoproduction off protons in ultra-peripheral p-Pb collisions at $\sqrt{s_{\rm NN}}=5.02$ TeV}",
    eprint = "1406.7819",
    archivePrefix = "arXiv",
    primaryClass = "nucl-ex",
    reportNumber = "CERN-PH-EP-2014-149",
    doi = "10.1103/PhysRevLett.113.232504",
    journal = "Phys. Rev. Lett.",
    volume = "113",
    number = "23",
    pages = "232504",
    year = "2014"
}

@article{Aktas:2005xu,
	author         = "Aktas, A. and others",
	title          = "{Elastic J/psi production at HERA}",
	collaboration  = "H1",
	journal        = "Eur. Phys. J.",
	volume         = "C46",
	year           = "2006",
	pages          = "585-603",
	doi            = "10.1140/epjc/s2006-02519-5",
	eprint         = "hep-ex/0510016",
	archivePrefix  = "arXiv",
	primaryClass   = "hep-ex",
	reportNumber   = "DESY-05-161",
	SLACcitation   = "%%CITATION = HEP-EX/0510016;%%"
}

@article{Breitweg:1997ed,
	author = "Breitweg, J. and others",
	collaboration = "ZEUS",
	title = "{Elastic and proton dissociative $\rho^0$ photoproduction at HERA}",
	eprint = "hep-ex/9712020",
	archivePrefix = "arXiv",
	reportNumber = "DESY-97-237",
	doi = "10.1007/s100520050136",
	journal = "Eur. Phys. J. C",
	volume = "2",
	pages = "247--267",
	year = "1998"
}

@article{Breitweg:1998ki,
	author = "Breitweg, J. and others",
	collaboration = "ZEUS",
	title = "{Measurement of elastic Upsilon photoproduction at HERA}",
	eprint = "hep-ex/9807020",
	archivePrefix = "arXiv",
	reportNumber = "DESY-98-089, ANL-HEP-PR-98-96",
	doi = "10.1016/S0370-2693(98)01081-8",
	journal = "Phys. Lett. B",
	volume = "437",
	pages = "432--444",
	year = "1998"
}

@article{Chekanov:2002xi,
	author         = "Chekanov, S. and others",
	title          = "{Exclusive photoproduction of J / psi mesons at HERA}",
	collaboration  = "ZEUS",
	journal        = "Eur. Phys. J.",
	volume         = "C24",
	year           = "2002",
	pages          = "345-360",
	doi            = "10.1007/s10052-002-0953-7",
	eprint         = "hep-ex/0201043",
	archivePrefix  = "arXiv",
	primaryClass   = "hep-ex",
	reportNumber   = "DESY-02-008",
	SLACcitation   = "%%CITATION = HEP-EX/0201043;%%"
}

@article{Chekanov:2007zr,
	author = "Chekanov, S. and others",
	collaboration = "ZEUS",
	title = "{Exclusive rho0 production in deep inelastic scattering at HERA}",
	eprint = "0708.1478",
	archivePrefix = "arXiv",
	primaryClass = "hep-ex",
	reportNumber = "DESY-07-118",
	doi = "10.1186/1754-0410-1-6",
	journal = "PMC Phys. A",
	volume = "1",
	pages = "6",
	year = "2007"
}

@article{Chekanov:2009zz,
	author = "Chekanov, S. and others",
	collaboration = "ZEUS",
	title = "{Exclusive photoproduction of upsilon mesons at HERA}",
	eprint = "0903.4205",
	archivePrefix = "arXiv",
	primaryClass = "hep-ex",
	reportNumber = "DESY-09-036",
	doi = "10.1016/j.physletb.2009.07.066",
	journal = "Phys. Lett. B",
	volume = "680",
	pages = "4--12",
	year = "2009"
}

@article{CMS:2018bbk,
    author = "Sirunyan, Albert M. and others",
    collaboration = "CMS",
    title = "{Measurement of exclusive $\Upsilon$ photoproduction from protons in pPb collisions at $\sqrt{s_\mathrm{NN}} =$ 5.02 TeV}",
    eprint = "1809.11080",
    archivePrefix = "arXiv",
    primaryClass = "hep-ex",
    reportNumber = "CMS-FSQ-13-009, CERN-EP-2018-225",
    doi = "10.1140/epjc/s10052-019-6774-8",
    journal = "Eur. Phys. J. C",
    volume = "79",
    number = "3",
    pages = "277",
    year = "2019",
    note = "[Erratum: Eur.Phys.J.C 82, 343 (2022)]"
}

@article{CMS:2019awk,
    author = "Sirunyan, Albert M and others",
    collaboration = "CMS",
    title = "{Measurement of exclusive $\rho(770)^0$ photoproduction in ultraperipheral pPb collisions at $\sqrt{s_\mathrm{NN}} =$ 5.02 TeV}",
    eprint = "1902.01339",
    archivePrefix = "arXiv",
    primaryClass = "hep-ex",
    reportNumber = "CMS-FSQ-16-007, CERN-EP-2018-285",
    doi = "10.1140/epjc/s10052-019-7202-9",
    journal = "Eur. Phys. J. C",
    volume = "79",
    number = "8",
    pages = "702",
    year = "2019"
}

@article{H1:2009wnw,
    author = "Aaron, F. D. and others",
    collaboration = "H1",
    title = "{Deeply Virtual Compton Scattering and its Beam Charge Asymmetry in e+- Collisions at HERA}",
    eprint = "0907.5289",
    archivePrefix = "arXiv",
    primaAaij:2014iearyClass = "hep-ex",
    reportNumber = "DESY-09-109, DESY09-109",
    doi = "10.1016/j.physletb.2009.10.035",
    journal = "Phys. Lett. B",
    volume = "681",
    pages = "391--399",
    year = "2009"
}

@article{H1:2020lzc,
	author = "Andreev, V. and others",
	collaboration = "H1",
	title = "{Measurement of exclusive $\pi^+ \pi ^-$ and $\rho^0$ meson photoproduction at HERA}",
	eprint = "2005.14471",
	archivePrefix = "arXiv",
	primaryClass = "hep-ex",
	reportNumber = "DESY-20-080",
	doi = "10.1140/epjc/s10052-020-08587-3",
	journal = "Eur. Phys. J. C",
	volume = "80",
	number = "12",
	pages = "1189",
	year = "2020"
}

@article{LHCb:2018rcm,
    author = "Aaij, Roel and others",
    collaboration = "LHCb",
    title = "{Central exclusive production of $J/\psi$ and $\psi(2S)$ mesons in $pp$ collisions at $\sqrt{s}=13~$TeV}",
    eprint = "1806.04079",
    archivePrefix = "arXiv",
    primaryClass = "hep-ex",
    reportNumber = "LHCB-PAPER-2018-011, LHCb-PAPER-2018-011, CERN-EP-2018-152",
    doi = "10.1007/JHEP10(2018)167",
    journal = "JHEP",
    volume = "10",
    pages = "167",
    year = "2018"
}

@article{ZEUS:1996zse,
    author = "Derrick, M. and others",
    collaboration = "ZEUS",
    title = "{Measurement of elastic omega photoproduction at HERA}",
    eprint = "hep-ex/9608010",
    archivePrefix = "arXiv",
    reportNumber = "DESY-96-159",
    doi = "10.1007/s002880050297",
    journal = "Z. Phys. C",
    volume = "73",
    pages = "73--84",
    year = "1996"
}

@article{ZEUS:2000swq,
    author = "Breitweg, J. and others",
    collaboration = "ZEUS",
    title = "{Measurement of exclusive omega electroproduction at HERA}",
    eprint = "hep-ex/0006013",
    archivePrefix = "arXiv",
    reportNumber = "DESY-00-084",
    doi = "10.1016/S0370-2693(00)00794-2",
    journal = "Phys. Lett. B",
    volume = "487",
    pages = "273--288",
    year = "2000"
}

@article{ZEUS:2005bhf,
    author = "Chekanov, S. and others",
    collaboration = "ZEUS",
    title = "{Exclusive electroproduction of phi mesons at HERA}",
    eprint = "hep-ex/0504010",
    archivePrefix = "arXiv",
    reportNumber = "DESY-05-038",
    doi = "10.1016/j.nuclphysb.2005.04.009",
    journal = "Nucl. Phys. B",
    volume = "718",
    pages = "3--31",
    year = "2005"
}

\end{document}